\documentclass[numberedappendix,apj,twocolumn]{emulateapj}

\newcommand{\bFilter}{$B_{435}$}
\newcommand{\vFilter}{$V_{606}$}
\newcommand{\iFilter}{$i_{775}$}
\newcommand{\zFilter}{$z_{850}$}
\newcommand{\yFilter}{$Y_{105}$}
\newcommand{\jhFilter}{$JH_{140}$}
\newcommand{\ymFilter}{$Y_{098}$}

\newcommand{\iwFilter}{$I_{814}$}

\newcommand{\jFilter}{$J_{125}$}

\newcommand{\hFilter}{$H_{160}$}

\newcommand{\msol}{$M_\odot$}

\shorttitle{Probing the Dawn of Galaxies at $z\sim9-12$}
\shortauthors{Oesch et al.}

\begin{document}

\title{Probing the Dawn of Galaxies at $z\sim9-12$: \\New Constraints from HUDF12/XDF and CANDELS Data
\altaffilmark{1}}

\altaffiltext{1}{Based on data obtained with the \textit{Hubble Space Telescope} operated by AURA, Inc. for NASA under contract NAS5-26555. }

\author{P. A. Oesch\altaffilmark{2,\dag},
R. J. Bouwens\altaffilmark{3}, 
G. D. Illingworth\altaffilmark{2}, 
I. Labb\'{e}\altaffilmark{3}, 
M. Franx\altaffilmark{3}, 
P. G. van Dokkum\altaffilmark{4},\\
M. Trenti\altaffilmark{5}, 
M. Stiavelli\altaffilmark{6},
V. Gonzalez\altaffilmark{7},
D. Magee\altaffilmark{2}
}

\altaffiltext{2}{UCO/Lick Observatory, University of California, Santa Cruz, 1156 High St, Santa Cruz, CA 95064; poesch@ucolick.org}
\altaffiltext{3}{Leiden Observatory, Leiden University, NL-2300 RA Leiden, Netherlands}
\altaffiltext{4}{Department of Astronomy, Yale University, New Haven, CT 06520}
\altaffiltext{5}{Institute of Astronomy, University of Cambridge, Madingley Road, Cambridge CB3 0HA, UK}
\altaffiltext{6}{Space Telescope Science Institute, 3700 San Martin Drive, Baltimore, MD 21218, USA}
\altaffiltext{7}{University of California, Riverside, 900 University Ave, Riverside, CA 92507, USA}
\altaffiltext{\dag}{Hubble Fellow}
%

%

\begin{abstract}

We present a comprehensive analysis of $z>8$ galaxies based on
ultra-deep WFC3/IR data. We constrain the evolution of the UV
luminosity function (LF) and luminosity densities from $z\sim11$ to $z\sim8$ by
exploiting all the WFC3/IR data over the Hubble Ultra-Deep Field from
the HUDF09 and the new HUDF12 program, in addition to the HUDF09
parallel field data, as well as wider area WFC3/IR imaging over
GOODS-South. Galaxies are selected based on the Lyman Break Technique
in three samples centered around $z\sim9$, $z\sim10$ and $z\sim11$, with seven $z\sim9$
galaxy candidates, and one each at $z\sim10$ and $z\sim11$. We confirm a new
$z\sim10$ candidate (with $z=9.8\pm0.6$) that was not convincingly identified
in our first $z\sim10$ sample. The deeper data over the HUDF confirms all
our previous $z\gtrsim7.5$ candidates as genuine high-redshift candidates,
and extends our samples to higher redshift and fainter limits ($H_{160}\sim29.8$ mag). 
We perform one of the first estimates of the $z\sim9$ UV LF and
improve our previous constraints at $z\sim10$. Extrapolating the lower
redshift UV LF evolution should have revealed 17 $z\sim9$ and 9 $z\sim10$
sources, i.e., a factor $\sim3\times$ and 9$\times$ larger than observed. The inferred
star-formation rate density (SFRD) in galaxies above 0.7 $M_\odot$yr$^{-1}$
decreases by $0.6\pm0.2$ dex from $z\sim8$ to $z\sim9$, in good agreement with
previous estimates. The low number of sources found at $z>8$ is
consistent with a very rapid build-up of galaxies across $z\sim10$ to $z\sim8$.
From a combination of all current measurements, we find a best
estimate of a factor 10$\times$ decrease in the SFRD from $z\sim8$ to $z\sim10$,
following $(1+z)^{-11.4\pm3.1}$. Our measurements thus confirm our
previous finding of an accelerated evolution beyond $z\sim8$, and signify
a rapid build-up of galaxies with $M_{UV}<-17.7$ mag within only $\sim200$ Myr from
$z\sim10$ to $z\sim8$, in the heart of cosmic reionization.

\end{abstract}


\section{Introduction}

The launch of the WFC3/IR camera in 2009 signified a major milestone in our
ability to observe galaxies within the cosmic reionization epoch at $z\gtrsim6$.
Thanks to its $\sim40$ times higher efficiency for detecting galaxies in the
near-infrared (NIR) compared to  previous cameras on the Hubble Space
Telescope (HST) we have pushed the observational frontier to within only
$\sim450$ Myr from the Big Bang. In its first year of operation WFC3/IR
resulted in the detection of $\sim130$ new galaxies at $z>6$ \citep[see e.g.][]{Bouwens11c}. Three years of
science operations of WFC3/IR and several deep extra-galactic surveys have
now resulted in a large sample of more than 200 galaxies in the
reionization epoch, primarily at $z\sim7$ and $z\sim8$ \citep[e.g.][]{Bouwens11c,Oesch12b,McLure12,Schenker12,Lorenzoni13,Bradley12,Yan11,Finkelstein12b,Grazian12}


From these samples it has become clear that the build-up of galaxies during
the first Gyr was a gradual process at $z<8$. The end of
reionization left no noticeable imprint on the galaxy population (at least down to the current limits of $M_{UV}\simeq-18$
-- corresponding to a star-formation rate of SFR$~\simeq1 M_\odot$yr$^{-1}$). 
The build-up of the UV luminosity function (LF) progresses smoothly
across the $z\sim6$ reionization boundary, following a constant trend all the
way from $z\sim8$ to $z\sim4$.  Galaxies typically become brighter by $\sim30-40\%$ 
per unit redshift accompanied by a proportional (but somewhat larger) increase in the average
star-formation rate of galaxies \citep[see e.g.][]{Smit12,Papovich11}. 


Given the large samples of galaxies discovered at $z\sim7-8$, the current
observational frontier is at $z\sim9$ and at earlier times. This is a
period when significant evolution of the galaxy UV LF is expected from models \citep[e.g.][]{Trenti10,Lacey11,Finlator11b}.
The observational evidence has been suggestive of a significant drop on the UV luminosity density (LD) at $z>8$, but
has not been conclusive \citep{Bouwens11a,Oesch12a}. As a result, the extent to which the luminosity
function and the star formation rate density are evolving at $z > 8$ has been
the subject of some debate \citep[see, e.g.,][]{Coe13,Zheng12}.

At these early epochs, current galaxy samples are still very small as HST is
approaching its limits. Initially, only one $z \sim 10$ galaxy candidate was
identified (UDFj-39546284), even in extremely deep WFC3/IR imaging of the
Hubble Ultra Deep Field (HUDF) as part of the HUDF09 survey \citep[][]{Bouwens11a}.  When combined with all the 
existing data over the Chandra deep field South (CDFS), this one source
suggested that the galaxy population is changing quickly, building up very
rapidly from $z\sim10$ to $z\sim8$.  In galaxies with SFR\ $>1 M_\odot$yr$^{-1}$ (equivalent to
$M_{UV}\simeq-18$ mag), the inferred UV luminosity density (LD) was found to increase by more than an order of magnitude 
in only $\sim200$ Myr from $z\sim10$ to $z\sim8$ \citep{Oesch12a}.
This is a factor $\sim5$ times larger than what would have been expected from a simple extrapolation
of the lower redshift trends of the UV LF evolution to $z\sim10$.

Several datasets have allowed us to improve these first constraints. The multi-cycle treasury program CLASH (PI: Postman)
 has provided four sources at $z\gtrsim9$ \citep{Zheng12,Coe13,Bouwens12CLASH}. In particular, the
detections of three $z\sim9$ galaxies around CLASH clusters by
\citet{Bouwens12CLASH} have provided a valuable estimate of the luminosity density in this
key redshift range.  Estimating volume densities from highly-magnified
sources found behind strong lensing clusters is challenging, involving systematic uncertainties due to the lensing model.  \citet{Bouwens12CLASH} used a novel technique of comparing
the $z\sim9$ source counts to those at $z\sim8$ in the same clusters
and obtained a good relative luminosity density estimate.  Since the $z \sim 8$ density
is well-established from the field \citep[e.g.][]{Oesch12b,Bradley12}, this gave a more robust measure than
trying to infer source densities directly using lensing models.  Interestingly, the
three $z \sim 9$ candidates from \citet{Bouwens12CLASH} are completely
consistent with the observed drop in the UV luminosity density and an accelerated evolution 
of the galaxy population that was previously seen at $z >8.5$ in the HUDF and CDFS by \citet{Oesch12a}.

Two other high redshift sources detected in the CLASH dataset \citep[one of which is in common with and proceeded the sample of][]{Bouwens12CLASH} have added to
the available constraints. \citet{Coe13} and \citet{Zheng12}
discovered two highly magnified $z \sim 10$  ($z\sim9.6$ and $z\sim10.7$) galaxies in the
analysis of the CLASH cluster data. The luminosity densities inferred from
these galaxies are somewhat higher, but are very uncertain. The large
errors on the luminosity density from these two detections encompass a wide
range of possible trends from $z \sim 8$ to earlier times.  However, as we show later in this
paper, taken together, the sources from the CLASH dataset along with the
latest sources and constraints from the HUDF/CDFS region, are consistent
with our earlier estimates of substantial accelerated change in the luminosity density
from $z\sim10$ to $z\sim8$.

Additional progress in exploring the galaxy population at $z>8$ has been made through gamma-ray burst (GRB) afterglow observations. The current record holder of an independently confirmed redshift measurement was achieved at $z\sim8.2$ for GRB090423 \citep{Tanvir09,Salvaterra09}, and GRB redshifts were photometrically measured out to $z=9.4$ \citep{Cucchiara11}. These measurements can provide additional constraints on the total star-formation rate density in the very early universe, since GRB rates are thought to be an unbiased tracer of the total star-formation rate density \citep[e.g.][]{Kistler09,Trenti12b,Robertson12}.

While the initial results are encouraging, it is clear that our understanding of the
galaxy population at $z > 8$ is still far from complete.  Given the very
small number of sources in each study, it is perhaps not surprising that
the UV luminosity density measurements at $z > 8$ are currently all within $1-2\sigma$ of each other. 
A next step forward in exploring the $z\gtrsim9$ universe can now be taken thanks to the 128 orbit HUDF12 campaign (PI: Ellis, GO12498). While the critical $H_{160}$ observations to discover $z\sim10$ galaxies only reach deeper by $\sim$0.2 mag compared to the previous HUDF09 image, the HUDF12 survey adds deep F140W ($JH_{140}$) imaging. This allows for Lyman Break Galaxy sample selections at $z\sim9$ and $z\sim11-12$ \citep[see also][]{Zheng12,Bouwens12CLASH,Coe13}.



\begin{deluxetable*}{lcccccccccc}
\tablecaption{The 5$\sigma$ Depths\tablenotemark{a} of the Observational Data Used in this Analysis\label{tab:data}}
\tablewidth{0 pt}
\tablecolumns{0}
\tablehead{\colhead{Field} & Area [arcmin$^2$] & \bFilter &\vFilter &\colhead{\iFilter} & \iwFilter &  \colhead{\zFilter}  &\colhead{$Y_{105}$} & \colhead{\jFilter} & \jhFilter & \colhead{\hFilter}  }
\startdata
HUDF12/XDF\tablenotemark{b,c} & 4.7 & 29.8  & 30.3  & 30.4 & 29.1  & 29.4  & 29.7  & 29.7 & 29.7\tablenotemark{*} & 29.8 \\
HUDF09-1 & 4.7 & --  & 29.5  & 29.3 & --  & 29.3  & 29.0  & 29.2 & -- & 29.0 \\
HUDF09-2 & 4.7 & 29.5  & 29.9  & 29.5 & --  & 29.2  & 29.0  & 29.2 & -- & 29.3 \\

ERS          & 41.3 & 28.4  & 28.7  & 28.2 & 28.5  & 28.0  & 27.8\tablenotemark{d}  & 28.2 & -- & 28.0 \\
GOODSS-Deep\tablenotemark{b} & 63.1 & 28.4  & 28.7  & 28.2 & 29.0  & 28.1  & 28.3  & 28.5 & -- & 28.3 \\
GOODSS-Wide & 41.9 & 28.4  & 28.7  & 28.2 & 28.5  & 28.0 & 27.5  & 27.7& --  & 27.5


\enddata

\tablenotetext{a}{Measured in circular apertures of 0\farcs25 radius.}
\tablenotetext{b}{Improved data relative to \citet{Oesch12a} for $z\sim10$ galaxy search.}
\tablenotetext{c}{A new version of the optical data is used here (the XDF data set; Illingworth et al.\ in prep.) compared to \citet{Beckwith06}, which results in an improvement in depth of $\sim0.1-0.2$ mag.}
\tablenotetext{d}{The ERS field was imaged with $Y_{098}$ rather than with $Y_{105}$.}
\tablenotetext{*}{Only the HUDF12/XDF field includes deep $JH_{140}$ imaging, which we require for $z\sim11$ searches and which significantly improves $z\sim9$ searches.}
%
\end{deluxetable*}

In a first analysis of their proprietary HUDF12 data, \citet{Ellis13} compiled a sample of six extremely faint $z\sim8.6-9.5$ galaxy candidates based on a photometric redshift technique. One of these sources was already in an earlier $z\sim7.2-8.8$ sample of \citet{Bouwens11c} based on the HUDF09 data set.
\citet{Ellis13} also re-analyzed our previously detected $z\sim10$ candidate UDFj-39546284 \citep{Bouwens11a,Oesch12a}. From the three years of WFC3/IR $H_{160}$ data, it is completely clear now that XDFjh-39546284 is a real source as it is significantly detected in all three
major sets of data taken in 2009, 2010 and 2012 \citep[][]{Bouwens12c}. 
Very surprisingly, however, UDFj-39546284 appears not to be detected in the new F140W image, indicating that this source is either a very extreme emission line galaxy at $z\sim2$ or that it lies at $z\sim12$ with the spectral break of the galaxy at $\sim$1.6$\micron$ \citep[see also][]{Brammer13}. 

With all the HUDF12 data publicly available, we can now extend our search for $z\gtrsim9$ galaxies to even deeper limits and to higher redshifts than previously possible. In this paper, we perform a search for $z\sim9-11$ galaxies over the HUDF based on the Lyman Break technique. 
This makes use of the fact that the hydrogen gas in the universe is essentially neutral at $z>6$, which results in near-complete absorption of rest-frame UV photons short-ward of the redshifted Ly$\alpha$ line. Star-forming galaxies at $z>6$ can therefore be selected as blue continuum sources which effectively disappear in shorter wavelength filters.
In Section \ref{sec:LBGvsZphot} we outline our reasons to use a Lyman Break
selection instead of photometric redshift selection, as is frequently adopted to identify very high-redshift galaxies in the literature, e.g.,  in
\citet{Ellis13}.

This paper is organized as follows: we start by describing the data used for this study in section \ref{sec:data} and define our source selection criteria in section \ref{sec:selection}, where we also present our $z\gtrsim9$ galaxy candidates. These are subsequently used to constrain the evolution of the UV luminous galaxy population out to $z\sim11$ in section \ref{sec:results}, where we present our results. In section \ref{sec:summary}, we summarize and discuss further possible progress in this field before JWST.

Throughout this paper, we will refer to the HST filters F435W, F606W, F775W, F850LP, F098M, F105W, F125W, F140W, F160W as \bFilter, \vFilter, \iFilter, \zFilter, \ymFilter, \yFilter, \jFilter, \jhFilter, \hFilter, respectively.
We adopt $\Omega_M=0.3, \Omega_\Lambda=0.7, H_0=70$ kms$^{-1}$Mpc$^{-1}$, i.e. $h=0.7$. Magnitudes are given in the AB system \citep{Oke83}.

\begin{figure}[tbp]
	\centering
	\includegraphics[width=\linewidth]{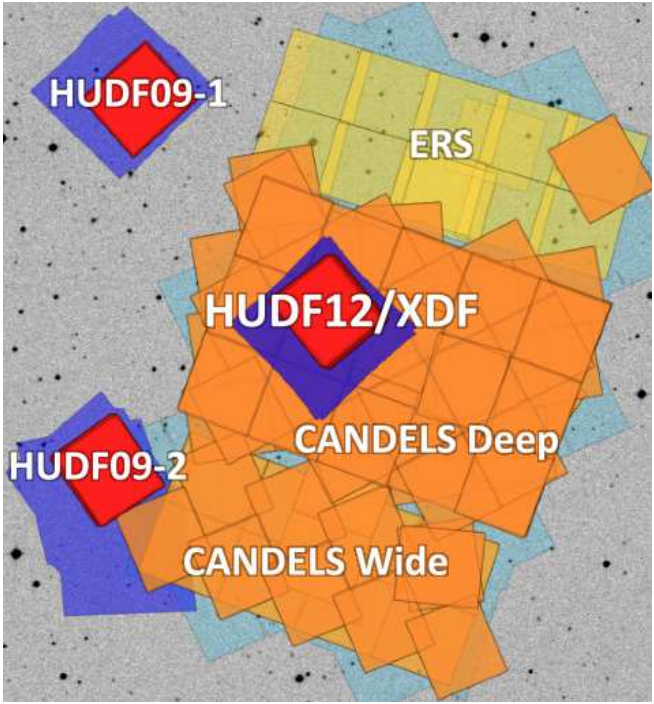}
  \caption{The WFC3/IR fields over the GOODS South area used in this analysis. The HUDF12/XDF field (dark red) contains the deepest optical and NIR data to date, which reach to $\sim30$ AB mag in several bands. The parallel fields HUDF09-1 and HUDF09-2 (also dark red) are only $0.5-0.8$ mag shallower. The wider area data covering the whole GOODS-S field are from the ERS (yellow) and the CANDELS programs (orange). All these fields include imaging in $Y_{105}$ (or $Y_{098}$ over the ERS), as well as $J_{125}$ and $H_{160}$, which makes it possible to search for $z\sim10$ galaxies. The HUDF12/XDF field is additionally covered by very deep $JH_{140}$ imaging, which we exploit to select $z\sim9$ Lyman Break galaxies and obtain some of the first limits on the galaxy population at $z\sim11$. }
	\label{fig:Fields}
\end{figure}

\section{The Data}
\label{sec:data}

The core dataset of this paper is the combination of ultra-deep ACS and
WFC3/IR imaging over the HUDF09/HUDF12/XDF field. We enhance this deep
dataset by using WFC3/IR and ACS data over both HUDF09 parallel fields, as
well as all CANDELS and ERS data over the GOODS-South field (see Figure \ref{fig:Fields}).
These datasets provide valuable constraints on the more luminous sources,
particularly by providing limits over a larger area than is covered by the
small HUDF09/HUDF12 field. All
these datasets include $J_{125}$ and $H_{160}$ imaging  in addition to deep,
multi-band optical ACS data, which allows for reliable $z\sim10$ galaxy selections (see Section  \ref{sec:z10}). 


All the WFC3/IR and ACS data are reduced following standard procedures. We subtract a super median image to improve the image flatness and we register to the GOODS ACS frames. For WFC3/IR data we mask pixels affected by persistence using the maps provided by STScI. The ACS data are corrected for charge transfer losses when necessary using the public code provided by STScI. All images are drizzled to a final pixel scale of $0\farcs06$, and the RMS maps are rescaled to match the actual flux fluctuations in circular apertures of 0\farcs35 diameter, dropped down on empty sky positions in the images. The spatial resolution of the data is $\sim0\farcs09$ and $\sim0\farcs16$ for the ACS and WFC3/IR data, respectively.  

The individual datasets used for our analysis are described in more detail in the following sections. They are furthermore summarized in Table \ref{tab:data} and are shown in Figure \ref{fig:Fields}.

\subsection{HUDF12/XDF Data}

The Hubble Ultra-Deep Field \citep[HUDF;][]{Beckwith06} was imaged with WFC3/IR as part of two large HST programs now. The HUDF09 \citep[PI: Illingworth;][]{Bouwens11c} provided one pointing (4.7 arcmin$^2$) of deep imaging in the three filters $Y_{105}$ (24 orbits), $J_{125}$ (34 orbits), and $H_{160}$ (53 orbits). These data were extended recently with the HUDF12 campaign \citep{Ellis13,Koekemoer12}, which imaged the HUDF further in $Y_{105}$ (72 orbits) and $H_{160}$ (26 orbits), and additionally added a deep exposure in $JH_{140}$ (30 orbits). These are the deepest NIR images ever taken, resulting in a final 5$\sigma$ depth of $H_{160} \sim 29.8$ mag (see also Table \ref{tab:data}). 

Since the acquisition of the original optical HUDF ACS data, several
programs have added deeper ACS coverage to this region, mainly as part of parallel imaging. We combined all the available ACS data over the HUDF, which allows us to
improve the backgrounds and also to push photometry limits deeper by $\sim0.1-0.2$ mag. These data, along with the matched WFC3/IR data from all
programs, will be released publicly as the eXtreme Deep Field (XDF) and are
discussed in more detail in Illingworth et al. (in preparation).

For longer wavelength constraints we also include the ultra-deep Spitzer/IRAC data from the 262 h IUDF program \citep[PI: Labb\'{e}; see also][]{Labbe12}, which reach to $\sim27$ mag AB (5$\sigma$ total) in both [3.6] and [4.5] channels. These data are extremely important for eliminating lower
redshift contaminating sources, particularly intermediate redshift dusty
and/or evolved galaxies (see section \ref{sec:dustycontamin}).

\subsection{HUDF09 Parallel Fields and GOODS-South}

In addition to the HUDF data, we also include the two additional deep parallel fields from the HUDF09 program, as well as all the WFC3/IR data over the GOODS South field. The latter were taken as part of the Early Release Science (ERS) program \citep{Windhorst11} and the multi-cycle treasury campaign CANDELS \citep[PI: Faber/Ferguson;][]{Grogin11,Koekemoer11}. These data were already used for a $z\sim10$ search in our previous analysis from \citet{Oesch12a}. We therefore refer the reader to that paper for a more detailed discussion.  However, since our previous analysis the acquisition of an additional 4 epochs of CANDELS DEEP data was completed, resulting in deeper data by about 0.2 mag. These are included now in this paper, which will allow us to further tighten our constraints on the $z\sim10$ LF.

In the optical, we make use of all ACS data taken over the GOODS South field, which includes additional imaging from supernova follow-up programs. These images reach $\sim0.1-0.3$ mag deeper than the v2.0 reductions of GOODS, in particular in the $z_{850}$-band. We also reduce and include all the $I_{814}$ data, which was taken over this field. 
By combining all these datasets we have produced what is the deepest optical image to
date over the GOODS-S field. Such deep optical data is very important for excluding lower redshift interlopers in LBG samples.

For constraints from Spitzer/IRAC, we use the public data from the GOODS campaign. These exposures are 23 h deep and reach to $\sim$26 mag (M. Dickinson et al. in prep).
All these fields are also outlined in Figure \ref{fig:Fields}.

\begin{figure}[tbp]
  \centering
	\includegraphics[width=\linewidth]{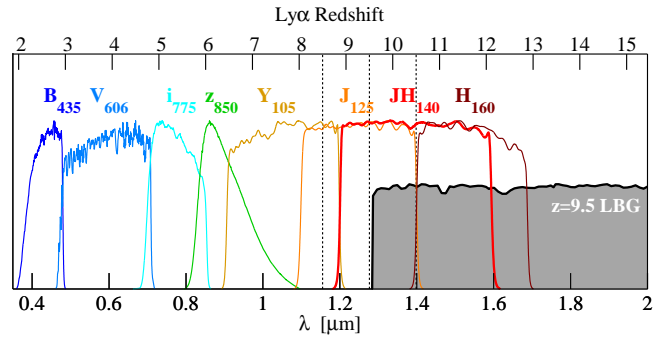} 
    \caption{The HST filter set over the HUDF12/XDF and the CANDELS fields together with a representative galaxy spectrum at $z=9.5$. Due to the high neutral fraction in the IGM, essentially all photons shortward of the redshifted Ly$\alpha$ line are absorbed for galaxies at $z>6$. This effect is used to select such high redshift sources based on broad-band photometry. The redshift of Ly$\alpha$ is indicated on the top axis. The vertical black dotted lines indicate the location of the break at $z=8.5,~9.5,$ and 10.5.  
    }
	\label{fig:filters}
\end{figure}

\begin{figure*}[tbp]
	\centering
	\includegraphics[width=0.45\linewidth]{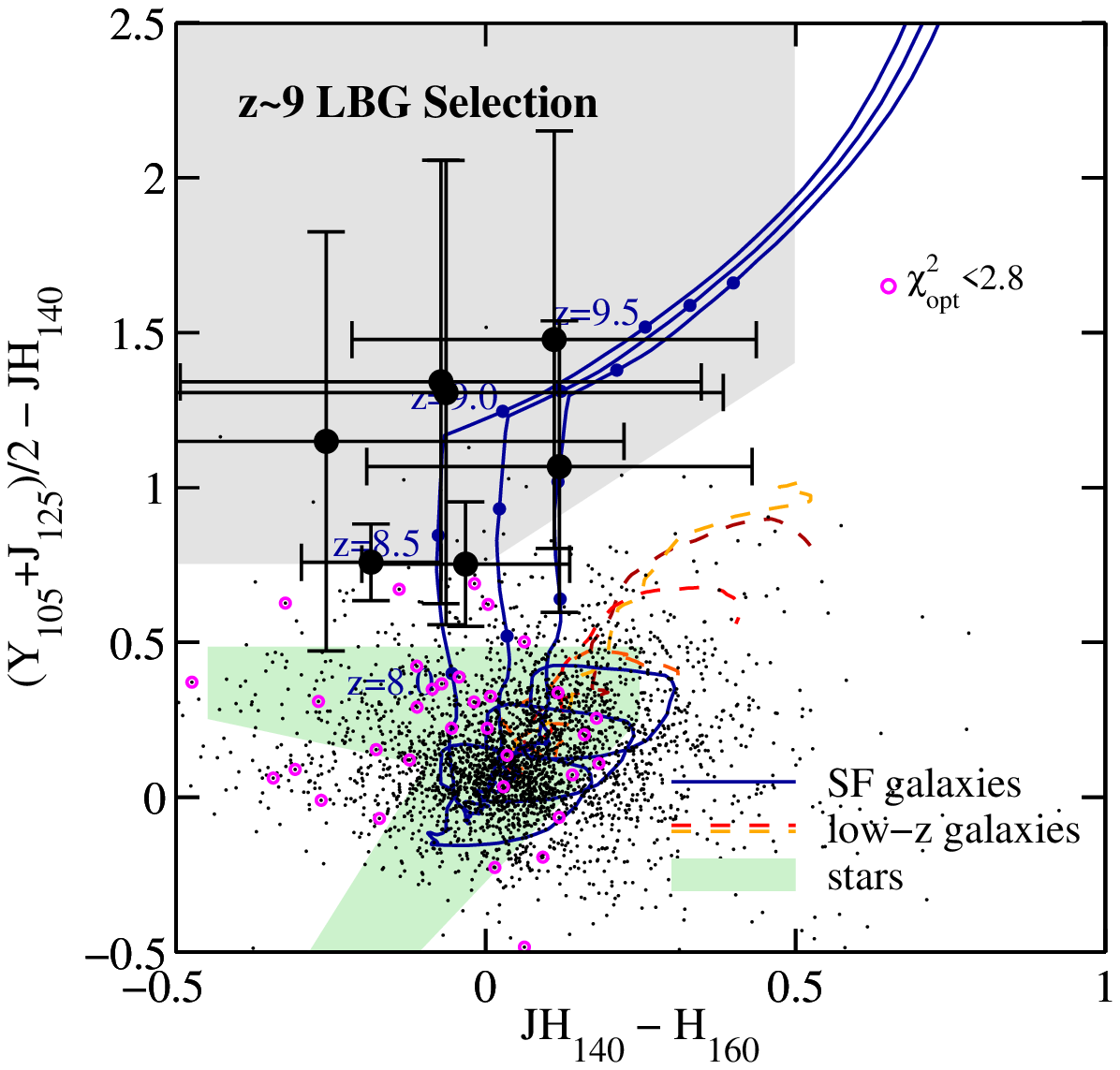} 
	\includegraphics[width=0.45\linewidth]{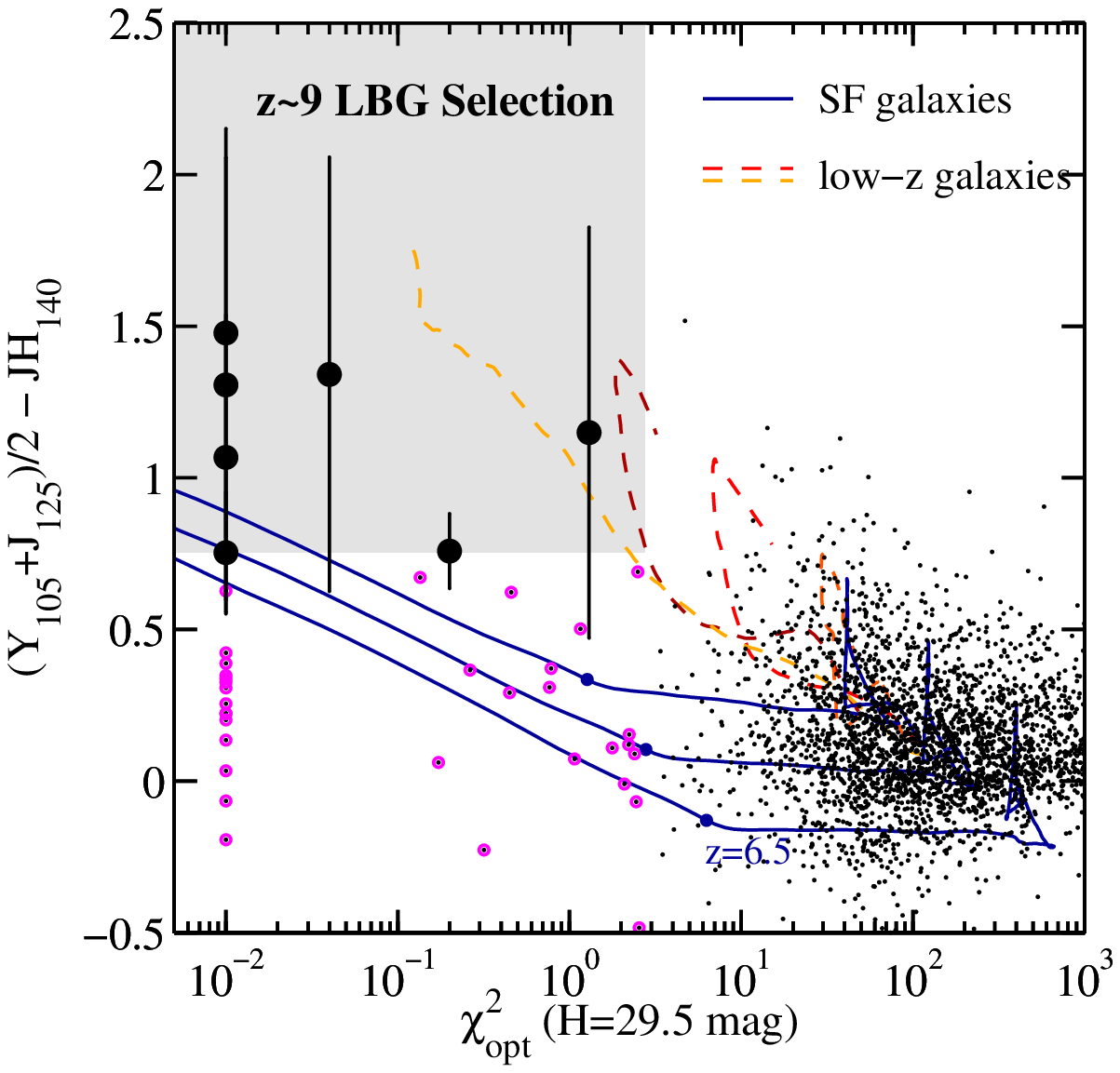} 
	\includegraphics[width=0.45\linewidth]{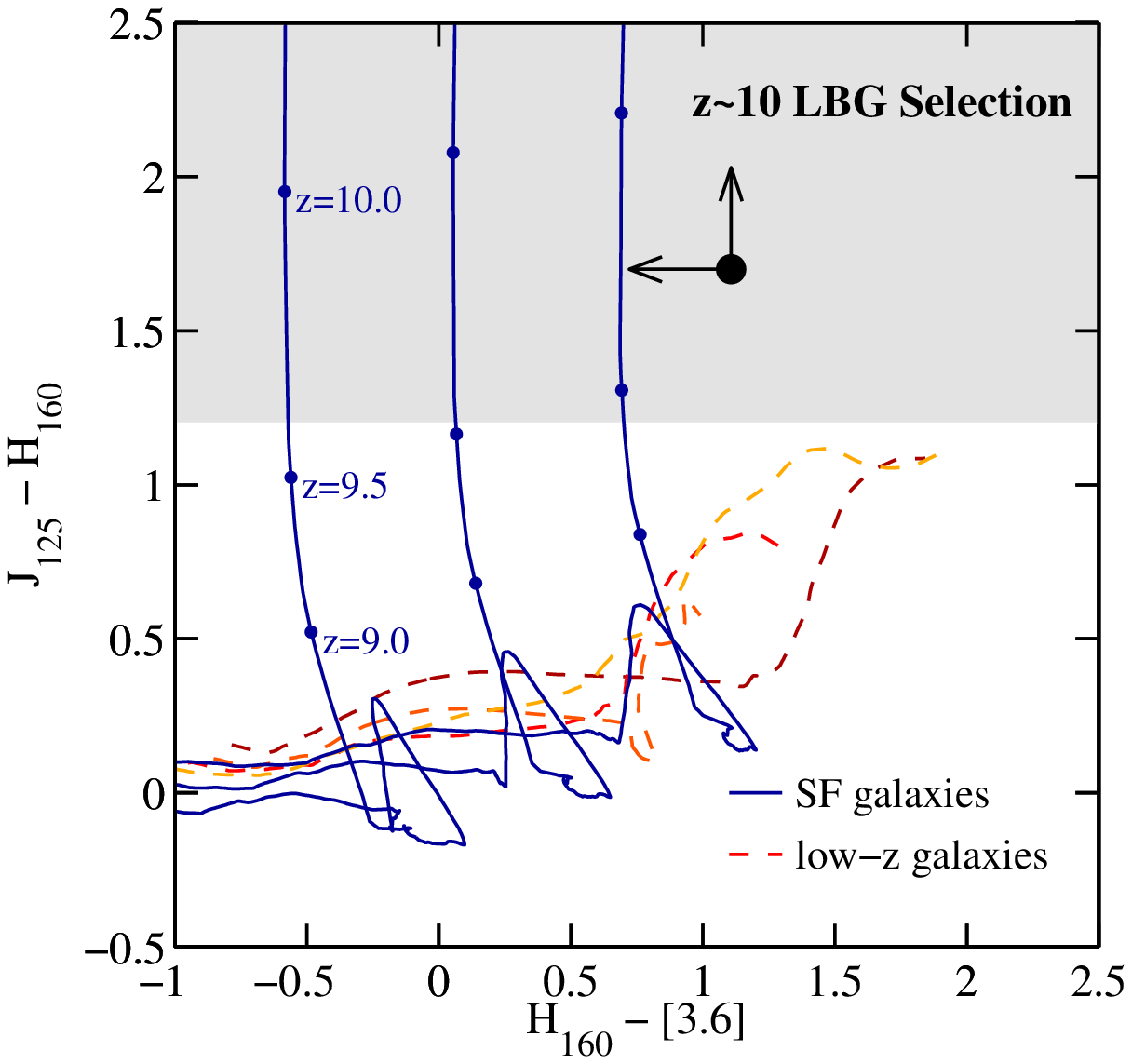}  
    \includegraphics[width=0.45\linewidth]{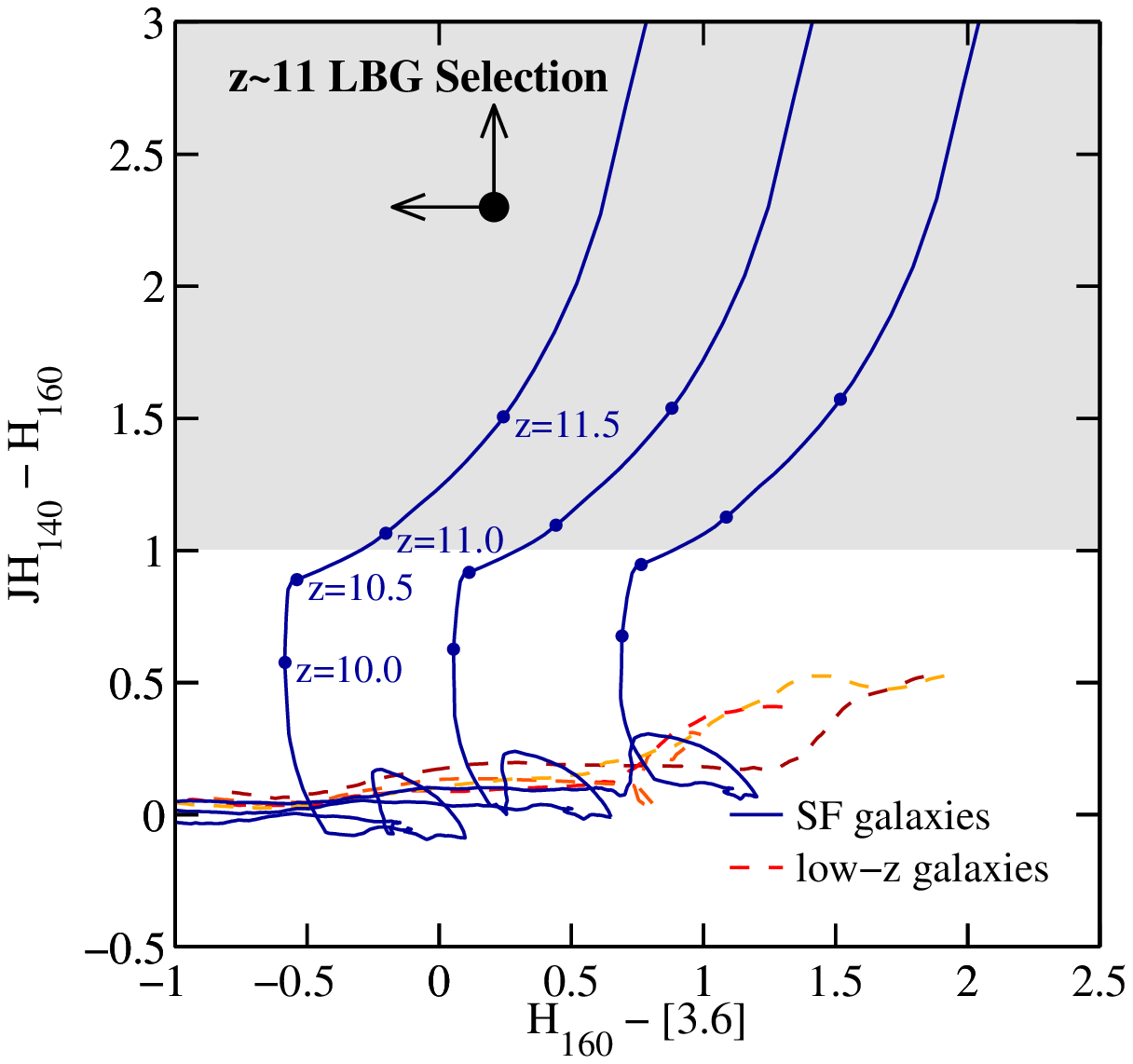}	
  \caption{The different color-color selections of our LBG samples. For the $z\sim9$ sample, we average the $Y_{105}$ and $J_{125}$ flux measurements to provide for a clean dropout selection which separates low redshift galaxies (dashed yellow to red lines) from star-forming sources at high redshift (blue). The different SF tracks assume a dust extinction of $E(B-V)=0,~0.15, ~0.30$ mag using a \citet{Calzetti00} reddening. In all panels, the color selections are indicated as light gray regions. The small black dots (in upper panels) show the full galaxy sample, while the large filled black circles represent the high-$z$ candidate sample. Small magenta circles in the upper panels represent galaxies with $\chi^2_{opt}<2.8$. As can be seen from their color distribution, the most likely sources to scatter into the $z\sim9$ selection are blue sources at just somewhat lower redshift ($z\sim7-8$) rather than intermediate redshift passive galaxies (yellow line at $JH_{140}-H_{160}>0.2$).  The upper right panel illustrates the $\chi^2_{opt}$ criterion for the $z\sim9$ selection, which guards our sample against $z<7$ sources.   Sources with $\chi^2_{opt}<0.01$ are limited at that value. As can be seen, the $z\sim9$ candidates lie in a quite unique region in the color-$\chi^2_{opt}$ plot, with only a few sources just outside our selection window.
  For the higher redshift samples (lower left and right panel), the primary selection criterion is the red color in the WFC3/IR filters. However, we additionally check for strong detections in both IRAC bands ([3.6] and [4.5]). The limits on colors are $1\sigma$.
  }
	\label{fig:colsel}
\end{figure*}

\section{Sample Selection}
\label{sec:selection}
%



Source catalogs are obtained with SExtractor \citep{Bertin96}, which is run
in dual image mode with a specific detection image, depending on the galaxy
sample we are interested in. For all samples at $z<10.5$, we use a $\chi^2$
detection image \citep{Szalay99} based on the $JH_{140}$ and $H_{160}$ bands. For $z>10$
$JH_{140}$-dropout selections we use the $H_{160}$-band for source
detection. 

All images are matched to the same PSF when performing photometry
measurements. Colors are based on small Kron apertures (Kron factor 1.2),
typically 0\farcs2 radius,
while magnitudes are derived from large apertures using the standard Kron
factor of 2.5, typically 0\farcs4 radius. An additional correction to total fluxes is performed based on
the encircled flux measurements of stars in the $H_{160}$ band to account for flux loss in the
PSF wings. This correction depends on the size of individual galaxies and
is typically $\sim0.2$ mag.

\subsection{Advantages of Lyman-Break over Photometric Redshift Selections}
\label{sec:LBGvsZphot}

In this paper, we adopt a Lyman Break galaxy selection to identify galaxies
at $z\gtrsim8.5$.  The major advantages of this approach over a
photometric redshift selected sample as is used, e.g., in \citet{Ellis13}
are simplicity and robustness. The Lyman Break technique provides a
straightforward and robust selection, which is easily reproducible by other teams \citep[e.g.][]{Schenker13}. 
Furthermore, the simplicity of the Lyman Break color-color criteria also allows for a straightforward estimate of the selection
volumes based on simulations.

In contrast, the photometric redshift likelihood functions
are heavily dependent on the assumed template set and even on the specific
photometric redshift code that is used. Additionally, the photometric
redshift likelihood functions depend on largely unknown priors which are
needed to account for the number density of intermediate redshift passive
or dusty galaxies. A particular problem is that most high-redshift
photometric analyses give equal weight to all templates at all redshift
(i.e.\ they adopt a flat prior). This includes faint galaxies with extreme
dust extinction at intermediate redshift or passive sources at $z>5$, which
are unlikely to be very abundant in reality.  

A further uncertainty, in particular for high redshift sources, is how undetected fluxes are treated in the fitting process. This can have significant influence on the lower redshift likelihood estimates.

Given all these advantages, we will
therefore select high-redshift galaxies using the Lyman Break technique and
we will determine their photometric redshifts a posteriori using standard
template fitting on this pre-selected sample of LBGs.


\begin{figure*}[tbp]
	\centering
	\includegraphics[width=0.83\linewidth]{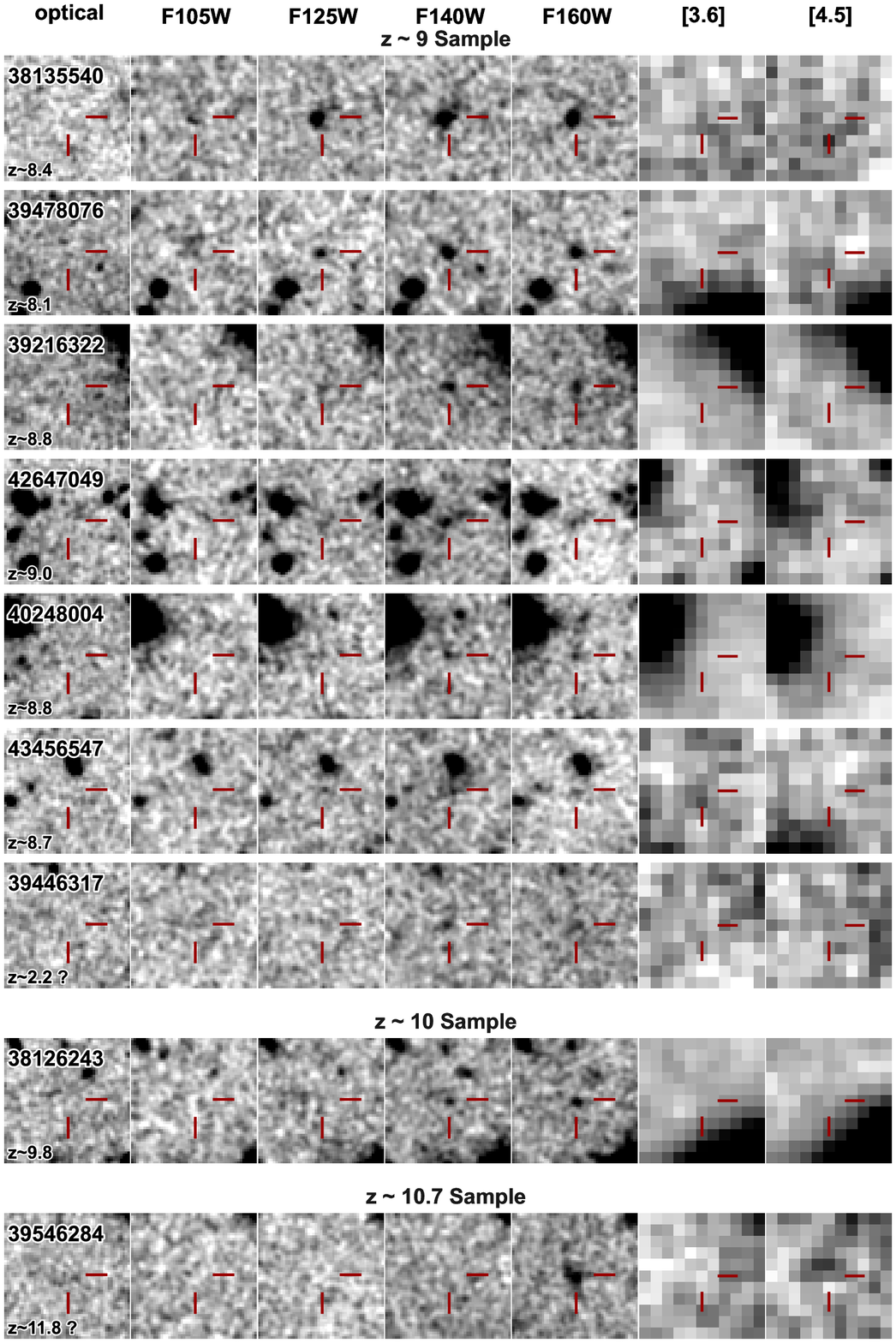}
  \caption{3\arcsec$\times$3\arcsec images of the $z>8$ galaxy candidates. From left to right, the images show, a stack of all optical bands, \yFilter, $JH_{140}$, \jFilter, \hFilter, IRAC [3.6], and [4.5]. The stamps are sorted by dropout sample. The approximate photometric redshift of each source is shown in the lower left corner of the optical stacked stamp (see also Table \ref{tab:phot}).}
	\label{fig:stampszgtr8}
\end{figure*}

\begin{figure*}[tbp]
	\centering
\includegraphics[width=0.95\linewidth]{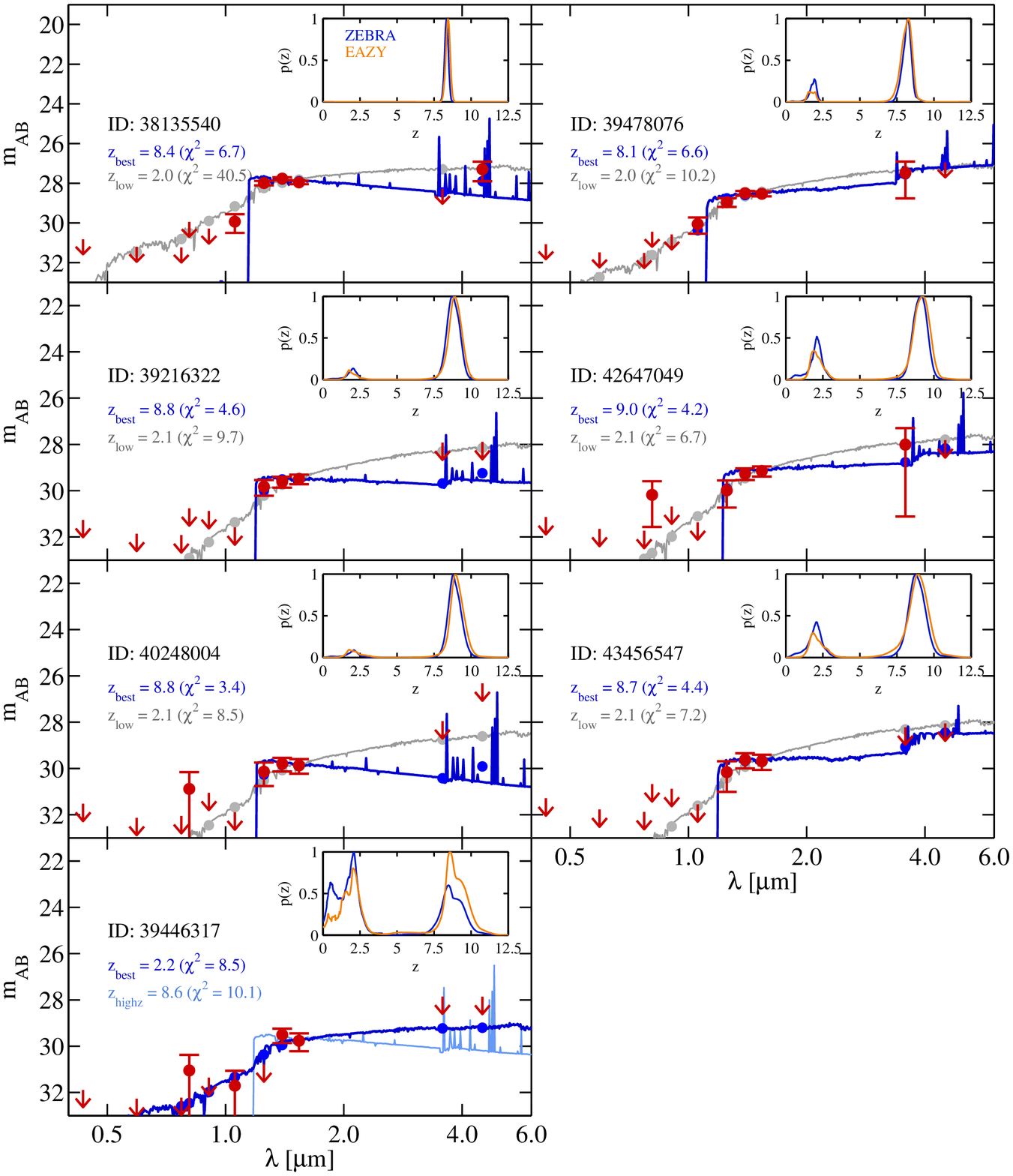}	  
  \caption{Spectral energy distribution fits to the fluxes of the $z\sim9$ $YJ$-dropout candidate galaxies in our sample. The magnitude measurements and upper limits ($1\sigma$) are shown in red. 
These also include self-consistent flux measurements in IRAC [3.6] and [4.5]. The uncertainties we derive
for the IRAC fluxes account for the uncertainties involved in removing 
contamination from neighboring sources, resulting in some variation
in the effective depth of the IRAC data.
The best-fit SEDs (blue) as well as the best low redshift solution (gray) are shown as solid lines. The SED magnitudes are indicated as filled circles. With the exception of source XDFyj-39446317 all sources have best-fit redshifts at $z\simeq8-9.5$. The low redshift solutions are evolved galaxies at $z\sim2$, for which the Lyman Break is confused with the Balmer/4000\AA\ break. As is evident, almost all other sources show a non-negligible secondary peak in their redshift likelihood distribution around $z\sim2$. 
Due to the best-fitting low redshift solution of XDFyj-39446317, we do not include this source in our analysis of the UV LF and SFRD. One such source of contamination is completely consistent with our expectation from photometric scatter simulations (see Section \ref{sec:scattersim}). }
	\label{fig:sedFits}
\end{figure*}

\subsection{$z\sim9$ Lyman Break Selection}

The addition of deep F140W imaging data over the HUDF gives us the ability
to select new samples of $z\sim9$ galaxies over that field.  As can be seen
in Figure \ref{fig:filters}, the absorption due to the inter-galactic
neutral hydrogen shifts in between the $Y_{105}$ and $J_{125}$ filters at
$z\gtrsim9$.  For a
robust Lyman Break selection, we thus combine the $Y_{105}$ and $J_{125}$ filter
fluxes in which galaxies start to disappear at $z\sim9$. Our adopted
selection criteria are:

\begin{eqnarray}
&	(Y_{105}+J_{125})/2 - JH_{140}>0.75  & 
\end{eqnarray}
\[
(Y_{105}+J_{125})/2 - JH_{140} > 0.75 + 1.3\times (JH_{140} - H_{160}) 
\]
\[
S/N(B_{435} \mathrm{ ~to~ } z_{850})<2  \quad \wedge \quad \chi^2_{opt}<2.8
\]

These criteria (shown in Figure \ref{fig:colsel}) are chosen to
select sources at $z\sim8.5-9.5$.  We additionally use a
$(J_{125}-H_{160})<1.2$ criterion to cleanly distinguish our $z\sim9$ and
$z\sim10$ samples (see next section).

We only include sources which are significantly detected in the $H_{160}$
and $JH_{140}$ images with at least 3$\sigma$ in each filter and with
$3.5\sigma$ in at least one of the two. As a cross-check we 
selected sources based on an inverse-variance weighted combination of the
$J_{125}$, $JH_{140}$, and $H_{160}$ images at $5\sigma$. Both selections
resulted in the same final sample of high-redshift sources, i.e. all selected candidates are $>5\sigma$ detections.

In addition to the color selections, we require sources to be
undetected at shorter wavelengths. In particular, we use $2\sigma$
non-detection criteria in all optical bands individually. Furthermore, we adopt an
optical pseudo-$\chi^2$ constraint. This is defined as 
$\chi^2_{opt} = \sum_i \mathrm{SGN}(f_i) (f_i/\sigma_i)^2$. The summation runs over all the
optical filter bands \bFilter, \vFilter, \iFilter, \iwFilter, and
\zFilter, and SGN is the sign function, i.e. $\mathrm{SGN}(x) = -1$ if
$x<0$ and $\mathrm{SGN}(x) = 1$ if $x>0$. This measure allows us to make
full use of all information in the optical data. We only consider galaxies with
$\chi^2_{opt}<2.8$. This cut reduces the contamination rate by a factor
$\sim3\times$, while it only reduces the selection volume of real sources
by 20\% \citep[see also][]{Oesch12b}.  This is a powerful tool for providing
source lists with low contamination rates (see also Section \ref{sec:scattersim}).

These selection criteria result in seven $z\sim9$ galaxy candidates in the HUDF12/XDF
dataset. These sources are listed in Table \ref{tab:phot} and their images are shown in
Figure \ref{fig:stampszgtr8}.  In Figure \ref{fig:sedFits}, we additionally show the spectral 
energy distribution (SED) fits and redshift likelihood functions for these sources. For comparison, these are
derived from two photometric redshift codes, ZEBRA \citep{Feldmann06,Oesch10c} as well as EAZY \citep{Brammer07}.
As is evident, the vast majority of sources does show a prominent peak at $z\sim8-9$ together with a
secondary, lower likelihood peak at $z\sim2$.
The best-fit photometric redshifts of these candidates range between $z=8.1-9.0$, with
the exception of one source (XDFyj-39446317), which has a ZEBRA photometric
redshift of only $z_{phot}=2.2$. However, using the EAZY code and template set the best-fit 
redshift is found at $z=8.6$. The photometric redshift likelihood function for this source is very wide using both codes. 

From photometric scatter simulations (see
section \ref{sec:scattersim}), we expect to find $0.9-1.1$ low redshift
contaminants in our $z\sim9$ sample due to photometric uncertainties.
Therefore, finding a LBG candidate with such a low redshift is not necessarily
unexpected. We will thus list it as possible candidate in Table \ref{tab:phot}.
However, we will exclude XDFyj-39446317 from our determination of the UV LF at
$z\sim9$.

\subsection{The $z\sim10$ Lyman Break Selection}
\label{sec:z10}

Galaxies at redshifts approaching $z\sim10$ start to disappear in the
$J_{125}$ filter. Following \citet{Bouwens11a} and \citet{Oesch12a}, we
select $z\sim10$ galaxies based on  very red $J_{125}-H_{160}$ colors
and we use Spitzer/IRAC photometry to guard this selection against
intermediate redshift extremely dusty and evolved galaxies in a second
step. This selection process also used $JH_{140}$ data when available (i.e. over the HUDF12/XDF field), and
was used for all the datasets shown in Figure \ref{fig:Fields}.

The HST selection criteria are:
\begin{equation}
	(J_{125}-H_{160})>1.2 \quad \wedge \quad (JH_{140}-H_{160})<1.0
\end{equation}
\[
S/N(B_{435} \mathrm{ ~to~ } Y_{105})<2\quad \wedge \quad \chi^2_{opt}<2.8
\]
in addition to at least 3$\sigma$ detections in both $H_{160}$ and $JH_{140}$ and  $>3.5\sigma$ in one of these. 
All sources in our final list also satisfy a $>5\sigma$ detection criterion in the combined $J_{125}+JH_{140}+H_{160}$ image.

The $JH_{140}-H_{160}$ color criterion was introduced to distinguish
$z\sim10$ from $z\sim11$ galaxies over the HUDF12/XDF field. The other
fields, which do not have deep $JH_{140}$ imaging, do not include this
criterion. We account for this difference in our analysis of the selection
functions (Section \ref{sec:selfun}).

When applying these selection criteria to the WFC3/IR+ACS data over
GOODS-S, we previously identified 16 galaxies which satisfied these
criteria. However, these are all extremely bright in the Spitzer/IRAC bands
and are even detectable in the shallow [5.8] and [8.0] channel data over
GOODS-S (having $H_{160}-[5.8]=2.4 - 4.0$ mag). These sources were therefore excluded from our $z\sim10$ analysis,
as their $H_{160}$ to IRAC colors were too red for a genuine $z\sim10$
galaxy. These are most likely $z\sim2-3$ galaxies with significant
extinction and possibly evolved stellar populations
\citep[see][]{Oesch12a}.

Even taking advantage of the deeper WFC3/IR data that became available over the CANDELS-South field subsequent 
to the \citet{Oesch12a} analysis, no new credible $z\sim10$ source could be found.
However, our selection revealed
three potential sources in the HUDF12/XDF data. Unfortunately, two of these
are very close to a bright, clumpy foreground galaxy. Their photometry is
therefore very uncertain, and it is unlikely that they are real
high-redshift sources. We nevertheless list these as potential sources in
Table \ref{tab:phot}.  However, we will not use them in the subsequent
analysis.

This leaves us with only one likely $z\sim10$ galaxy candidate in all the
fields we have analyzed here. This is XDFj-38126243, which we had previously
identified in the first-epoch data of the HUDF09 as a potential $z\sim10$ candidate \citep{Bouwens11a}.
However, it was not detected at a significant enough level in the
subsequent second epoch $H_{160}$ data to indicate at high confidence that
it was real. As a result it was not included in
our final sample of $z\sim10$ sources from the HUDF09 data. 

\begin{figure}[tbp]
	\centering
		\includegraphics[width=0.9\linewidth]{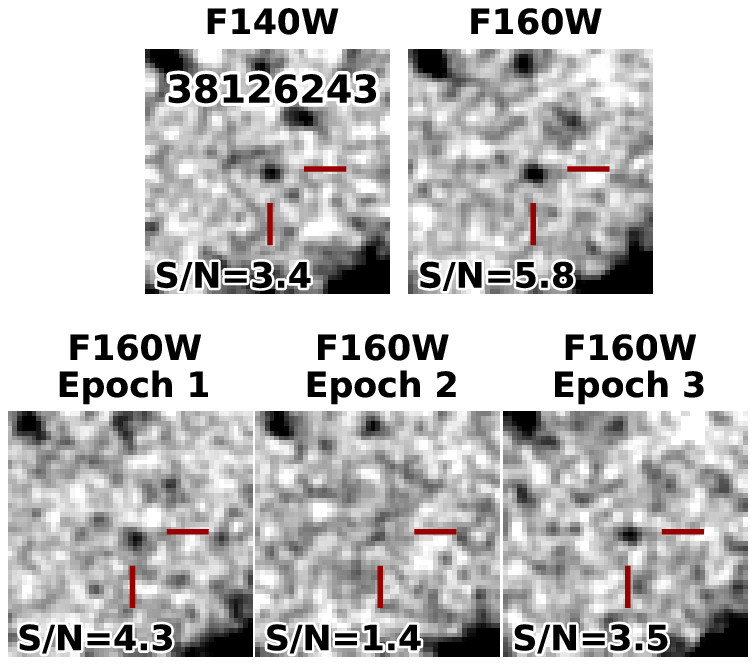}  
\includegraphics[width=\linewidth]{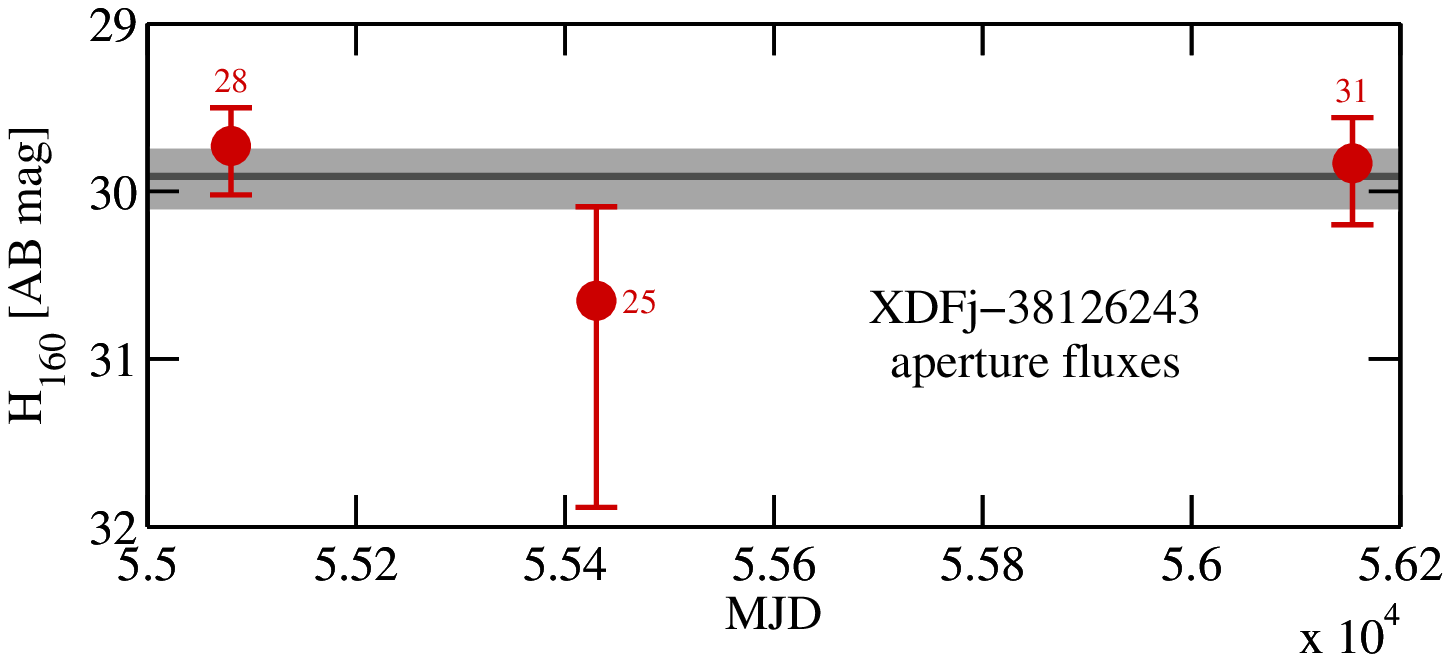} 
  \caption{\textit{Top --} Stamps (3\arcsec\ by 3\arcsec) of the $z\sim10$ candidate XDFj-38126243 -- which is likely the
highest-redshift source over the HUDF12/XDF. The top two stamps are the $JH_{140}$ and $H_{160}$ observations. The $H_{160}$ stack is also split in observations at three different epochs, shown in the bottom three stamps. `Epoch 1' corresponds to the HUDF09 year 1 data (28 orbits), `Epoch 2' are the HUDF09 year 2 data (25 orbits), and `Epoch 3' are the remaining 31 orbits from the HUDF12 and CANDELS observations. The S/N in each band is listed in the lower left. This source was initially selected as high-redshift candidate after the first year HUDF09 data \citep{Bouwens11a}.
However, as can be seen, the galaxy was only very weakly detected (1.4$\sigma$) in the HUDF09 year 2 data. Nevertheless, the source is clearly visible at $\gtrsim3.5\sigma$ in all other epochs, as well as in the $JH_{140}$ data (only taken from the HUDF12 program). The source is therefore clearly real. The lower signal detection in the `Epoch 2' data is still consistent with the expectation from a Gaussian noise distribution.
  \textit{Bottom --} The $H_{160}$ magnitude measurement for the three different epochs. The number of orbits going into each image is indicated close to each datapoint. The fluxes are measured in a circular aperture of 0\farcs35 diameter. The magnitude from the total 84-orbit $H_{160}$ image is indicated by the gray line, with errorbars represented by the filled gray area. 
The flux measurements are consistent with no variability in this source ($\chi^2 = 2.6$). However, an AGN contribution to its UV flux can not be excluded.}
	\label{fig:epochStamps}
\end{figure}

The source XDFj-38126243 is
now clearly detected both in the new $H_{160}$ and in the $JH_{140}$ data
from the HUDF12 survey, which clearly confirms its reality. This is demonstrated in
Figure \ref{fig:epochStamps}. As can also be seen from that figure, the
source is extremely compact, consistent with being a point source.  
We can therefore not exclude that this source is powered by an AGN, which could also explain the possible variability over a timescale of 1 year (see lower panel of Fig. \ref{fig:epochStamps}).
However, the low flux measurement in the second-year HUDF09 data is still consistent with expectations from Gaussian noise. Taken together, the flux measurements of all three epochs are consistent with the source showing no time variability ($\chi^2 = 2.6$).



The source is not detected in our ultra-deep IRAC data, and its colors place it at a photometric redshift of $9.8^{+0.6}_{-0.6}$ (see also Table \ref{tab:phot} and Figure \ref{fig:SEDfitZ10}).

For completeness, we also list three additional potential candidates in the appendix in Table \ref{tab:additional}. These sources  lie very close to bright foreground galaxies. Formally, they show colors consistent with being at very high redshift. However, they are significantly blended with their neighbors, such that the fluxes of these sources can not be accurately measured with SExtractor, without a more sophisticated neighbor subtraction technique. Additionally, the close proximity to very bright foreground sources casts significant doubt on the reality of these sources, and we therefore do not include these in our analysis.

\begin{figure}[tbp]
	\centering
\includegraphics[width=\linewidth]{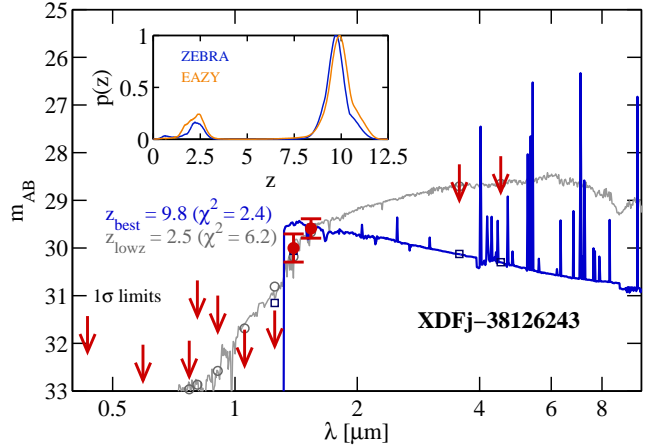}	
  \caption{Spectral energy distribution (SED) fit to the fluxes of the galaxy XDFj-38126243. The best-fit redshift is found at $z=9.8$, which is represented by the blue line. The gray SED corresponds to the best low-redshift solution at $z=2.5$. Open blue squares and gray circles show the expected magnitudes of these SEDs. 1$\sigma$ upper limits to undetected fluxes are shown as red arrows.  The inset shows the redshift likelihood function as estimated with ZEBRA (blue) and EAZY (orange). Both photometric redshift codes consistently find a prominent peak at $z=9.8$ and a lower-significance peak at $z\sim2.5$.}
	\label{fig:SEDfitZ10}
\end{figure}

\subsection{$z\sim10.7$ Lyman Break Selection}

The addition of $JH_{140}$ imaging from the HUDF12 data also allows for the
selection of $z\gtrsim10.5$ galaxies based on a red $JH_{140}-H_{160}$
color, as the IGM absorption shifts through the center of the $JH_{140}$
filter (see Figure \ref{fig:colsel}). We therefore search for galaxies in the
HUDF12/XDF data satisfying the following: 

\begin{equation}
	(JH_{140}-H_{160})>1.0
\end{equation}
\[
	 S/N(B_{435} \mathrm{ ~to~ } J_{125})<2 \quad \wedge \quad \chi^2_{opt}<2.8
\]

In order to ensure the reality of sources in this single-band detection sample we require $>5\sigma$ detections in $H_{160}$. 

Only one source satisfies these criteria: XDFj-39546284. This is our
previous highest redshift candidate from the HUDF09 data
\citep[see][]{Bouwens11a}.  Surprisingly, with $JH_{140}-H_{160}>2.3$ it has an extremely
red color in these largely overlapping filters. It is by far the reddest
source in the HUDF12/XDF data, and its rather dramatic color raises some
questions. 

At face value, the extreme color, together with the
non-detection in the optical data and in our deep IRAC imaging,
results in a best-fit redshift of $z=11.8\pm0.3$ for this source. However,
such a high redshift would imply that the source is $\sim20\times$ brighter
than expected for $z\sim12$ sources at the same number density
\citep[see][]{Bouwens12c}. While a strong Ly$\alpha$ emission line could
reduce the remarkably high continuum brightness, the lack of Ly$\alpha$
seen in galaxies within the reionization epoch at $z\gtrsim6$, indicates that the high fraction of neutral
hydrogen in the universe at early times absorbs the majority of Ly$\alpha$
photons of these galaxies \citep[e.g.][]{Schenker12,Pentericci11,Caruana12,Bunker13}.  Seeing strong Ly$\alpha$ emission from a source
at $z\sim12$ during the early phase of reionization is particularly
unexpected (although perhaps not impossible).

Alternatively, the source could be a $z\sim2$ extreme line emitter. An
emission line at $\sim1.6$\micron\ would have to produce the majority of its
$H_{160}$ flux, which would require extreme equivalent widths. A possible example of such a source 
is presented in \citet{Brammer13}.
For more extensive
discussions of these alternative options for this source see \citet{Brammer13} and \citet{Bouwens12c}.

Given the uncertain nature of this source, we will treat its detection as
an upper limit of $\leq1$ source in the following analysis, and we will
only derive upper limits on the luminosity and star-formation rate
densities at $z\sim10.7$ from this single-source sample.

\subsection{Sources of Sample Contamination}
\label{sec:contamin}

In the following sections we discuss possible contamination of our $z>8$ LBG samples.

\subsubsection{Dusty and Evolved Galaxies}
\label{sec:dustycontamin}

As already pointed out in \citet{Oesch12a}, galaxies with strong Balmer
breaks, or with high dust obscuration are a potential source of contamination
for $z>9$ galaxy searches. In particular, in fields with limited depth in
the WFC3/IR and optical data, such extremely red sources can remain
undetected shortward of $H_{160}$ and can thus satisfy the HST selection
criteria. Fortunately, with the availability of Spitzer/IRAC over all the
search fields in this study, such sources can readily be excluded from the
samples based on a $H_{160}-[3.6]<2$ color criterion. 

As shown in \citet{Oesch12a}, the CANDELS data contains 16 intermediate
brightness sources which satisfy our $z\sim10$ $J_{125}$-dropout
selections.  However, these could all be excluded based on the IRAC
constraints. They are all found at $H_{160}\sim24-26$ mag, which
suggests that such red, lower redshift galaxies show a peaked LF. Therefore,
it is expected that they would be much less of a problem as contaminants
in our fainter samples. 

Given this expectation, it is particularly interesting that we actually did
not find the lower luminosity counterparts of such sources in any of the
three deep fields, even though our IRAC data are sensitive enough  
thanks to the IUDF program. Although the survey volume of our
deep data is limited, this further suggests that such red galaxies are
indeed very rare at lower luminosities.  It remains an open and interesting
question as to the nature and redshift of these red $H_{160}\sim24-26$ mag
galaxies (the redshift is expected to be low, i.e. $z\sim1-3$, but exactly over what redshift
range they are seen is still quite uncertain).

\subsubsection{Photometric Scatter of Low-$z$ Sources}
\label{sec:scattersim}

After excluding contamination from intermediate redshift, red galaxies, the
next most important source of contamination is photometric scatter.
Photometric scatter can cause faint, low-redshift sources to have colors
and magnitudes such that they would be selected in our sample. We estimate the
magnitude of this effect with simulations using real galaxies based on our photometric
catalogs. 

In particular, we select all sources with $H_{160}$ magnitudes in the range
24 to 25 and we rescale their fluxes and apply the appropriate amount of  
photometric scatter as observed for real sources at fainter luminosities.
We then apply our selection criteria to these simulated catalogs in order
to estimate the contamination fraction. This is repeated 5000 times, which
results in reliably measured contamination fractions.

As expected, contamination due to photometric scatter is most significant
at the faint end of our sample. The above simulations show that we do not expect to see any
contaminant at $>1$ mag from the detection limit. In the HUDF12/XDF
$z\sim9$ galaxy sample, we find that 0.9 contaminants are expected per
simulation. Given that we find 7 sources, this signifies a $\sim12\%$
contamination fraction.  Note that this would have been a factor 3$\times$
higher (2.6 contaminants expected) had we not included our optical
$\chi^2$ measurement.  Again this shows clearly the power of having deep
shorter wavelength data, and the effectiveness with which data over a range
of wavelengths can be used. 

For the higher redshift samples, we estimate 0.2 and $<0.1$ contaminants in
the HUDF12/XDF LBG selection at $z\sim10$ and $z\sim10.7$, respectively,
from analogous simulations.  Overall, our extensive simulations show that the
contamination due to photometric scatter is thus expected to be
$\lesssim20\%$ for all these samples.

As an additional test of the contamination rate in our samples, we can use the 
best-fitting low-redshift SEDs for our $z\sim9$ candidates in order to estimate with
what probability such types of galaxies would be selected as LBGs.
In particular, we use the expected magnitudes of the $z\sim2$ SED fits shown in 
Figure \ref{fig:sedFits}, perturb these with the appropriate photometric scatter, and
apply our $z\sim9$ LBG selection. We repeat this simulation 10$^6$ times for each of our $z\sim9$ galaxy candidates, 
which allows us to estimate the probability for our $z\sim9$ sample to contain a certain number of contaminants.
We find that at 65\% confidence, our sample contains zero or one contaminant, while $\leq2$ contaminants are found in
90\% of the realizations. Finally, the chance that the majority (i.e., $>3$) of these $z\sim9$ candidates lie at $z\sim2$ is estimated 
to be $<1.5\%$.

Note that the average number of contaminants per realization is found to be 1.1, i.e., very similar
to our previous estimate of 0.9 contaminants based on using the real, bright galaxy population.  
It should be noted that the SED-based test makes no assumptions about the relative abundance of faint, star-forming $z>8$ galaxies and
the possible intermediate redshift passive sources at the same observed magnitude ($\sim29$ mag AB). Both observationally and theoretically, the number density of low-mass, passive galaxies at $z\gtrsim2$ is still very poorly understood, making it very difficult to gauge the contamination rates for our $z>8$ samples. In particular, these passive $z\sim2$ galaxies would need to have only $1-3\times10^8$ $M_\odot$ and $M_B = -15$ to $-16$ mag.

Nevertheless, these tests show that while we can not completely exclude intermediate redshift contamination in our samples, the majority of our 
candidates are clearly expected to lie at $z>8$. As we noted earlier, to account for this contamination we exclude the
$z\sim9$ candidate XDFyj-39446317 from the subsequent analysis, in agreement with its ambiguous photometric redshift at $z=2.2$ or $z=8.6$.

\subsubsection{Additional Sources of Contamination}

Galactic dwarf stars can be a significant concern for $z\sim7-8$ galaxy
samples, due to strong absorption features in their atmospheres, which
causes their intrinsic colors to overlap with the high-redshift galaxy selection criteria. However, this is not
as much of a concern at $z>9$. Stellar spectra are significantly bluer in
our selection colors than high-redshift galaxies.   This can be seen in
Figure \ref{fig:colsel}, where we plot the location of the stellar sequence
including M, L, and T dwarfs.  Stars with intrinsically red colors are therefore not expected to be a
significant contaminant in our samples. The only possibility for such stars to contaminate our selection is 
due to photometric scatter, which we implicitly accounted for in our photometric simulations in the previous section.

Additionally, we can exclude contamination by supernovae. We verified that
all galaxies in our sample are detected at statistically-consistent S/N
levels in the images taken over a time baseline of about 3yr as part of the
HUDF09 and HUDF12 campaigns. 

For a more extensive review of possible contamination in $z\gtrsim8$ samples, see
also \citet{Coe13} and \citet{Bouwens11c}.

\begin{figure}[tbp]
	\centering
	\includegraphics[width=\linewidth]{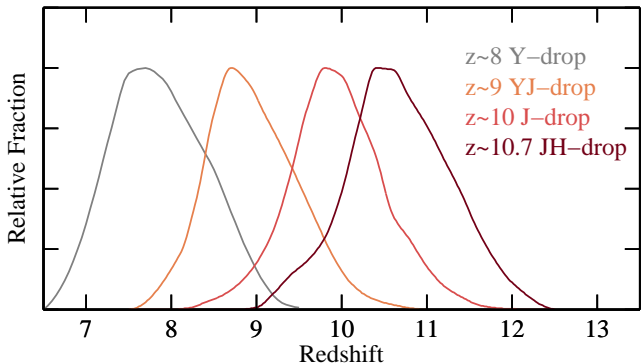}  
  \caption{The redshift selection function for different LBG samples. The new selections used in this paper nicely extend the lower redshift $Y$-dropout samples. The new $YJ$-dropout selection has a mean redshift $z=9.0$, while the $J$-dropout sample is expected to lie at a mean  $z=10.0$. Although the $JH_{140}-H_{160}>1$ color is only satisfied for $z>11$ galaxies, the $JH_{140}$-dropout sample extends to significantly lower redshift, and peaks only at $z\sim10.6$. This is mainly due to photometric scatter and due to the relatively slow change in $JH_{140}-H_{160}$ color from $z\sim10-11$ (see Figure \ref{fig:colsel}).   }
	\label{fig:zSelection}
\end{figure}

\subsection{Redshift Selection Functions}
\label{sec:selfun}

The expected redshift distributions of our LBG samples are estimated based on extensive simulations of artificial galaxies inserted in the real data which are then re-selected in the same manner as the original sources \citep[see also][]{Oesch07,Oesch12a}. In particular, we estimate the completeness $C(m)$ and selection probabilities $S(z,m)$ as a function of $H_{160}$ magnitude $m$ and redshift $z$. 

Following \citet{Bouwens03}, we use the profiles of $z\sim4$ LBGs from the HUDF and GOODS observations as templates for these simulations. The images of these sources are scaled to account for the difference in angular diameter distance as well as a size scaling of $(1+z)^{-1}$. The latter is motivated by observational trends of LBG sizes with redshift across $z\sim3-8$ \citep[see e.g.][]{Ferguson04,Bouwens04a,Oesch10b,Ono12}. 

The colors of the simulated galaxy population are chosen to follow a distribution of UV continuum slopes with $\beta=-2.4\pm0.4$ \citep[see e.g.][]{Bouwens09b,Bouwens10b,Finkelstein10,Stanway05}, and are modulated by the IGM absorption model of \citet{Madau95} over a range of redshifts $z=8$ to $z=13$. $10,000$ galaxies are simulated for each redshift bin in steps of $dz=0.2$, which allows for a reliable estimate of the completeness and selection probability taking into account the dispersion between input and output magnitudes. 

The result of these simulations enables us to compute the redshift distribution of galaxies after assuming a LF $\phi(M)$. 
\begin{equation}
p(z) = dN/dz = \int dm \frac{dV}{dz} S(m,z)C(m)\phi(M[m,z])
\end{equation}
We assume a baseline UV LF evolution with $\alpha=-1.73$ and $M_*(z) = -20.29 + 0.33 \times (z - 6)$, consistent with the trends found across $z\sim4$ to $z\sim8$ \citep{Bouwens11c}. The normalization is not relevant for the relative distributions. For the K-correction in the conversion from absolute to observed magnitude, we use a 100 Myr old, star-forming template of \citet{Bruzual03}.

The results of this calculation are shown in Figure \ref{fig:zSelection}. It is clear that the redshift selection functions are significantly wider than the target redshift range based on the simple LBG color tracks shown in Figure \ref{fig:colsel}. The reason for this is simply photometric scatter. The mean redshift (and the width) of our samples are 9.0 ($\pm0.5$), 10.0 ($\pm0.5$), and 10.7 ($\pm0.6$) for the $YJ$-, $J$- and $JH$-dropout samples, respectively.

\begin{figure}[tbp]
	\centering
   \includegraphics[width=0.7\linewidth]{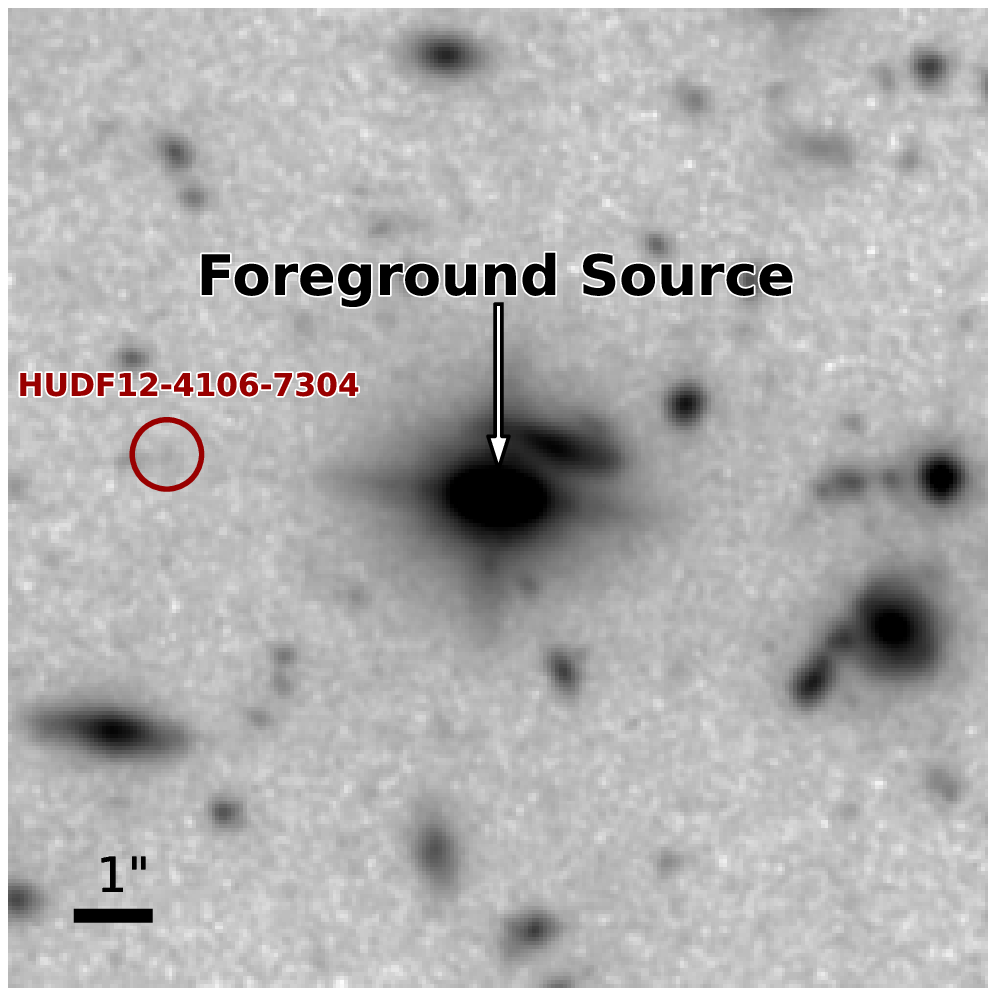}    
    \includegraphics[width=0.9\linewidth]{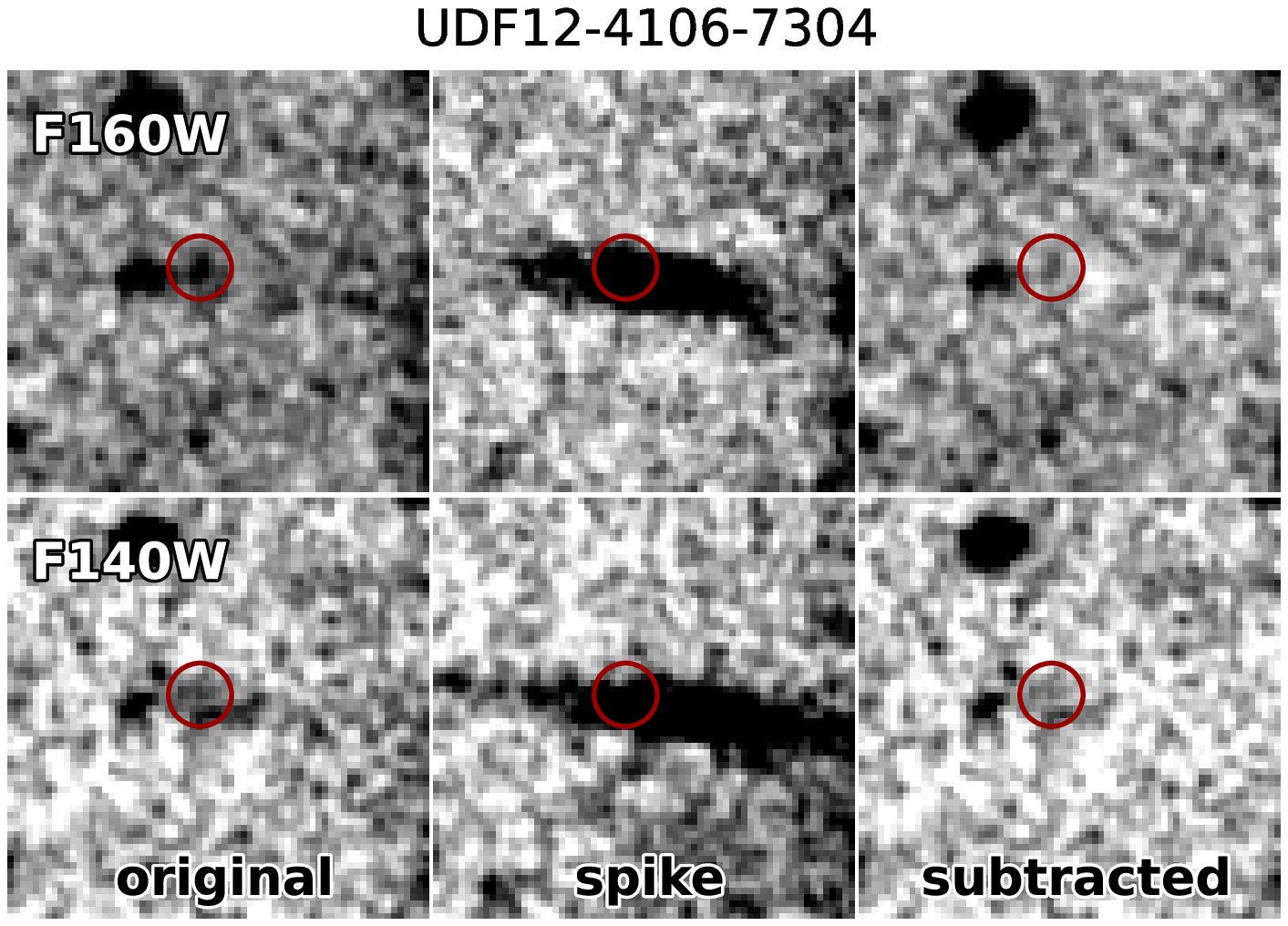}
  \caption{ The flux of the $z\sim9.5$ candidate HUDF12-4106-7304 of \citet{Ellis13} appears to be significantly boosted by a diffraction spike.
 \textit{Top --} A 15\arcsec view around a the bright foreground neighbor of the source HUDF12-4106-7304. The source clearly lies exactly along the direction of the diffraction spike caused by the bright neighbor. 
  \textit{Bottom --}  
   Stamps (4\arcsec\ by 4\arcsec) of the $z\sim9.5$ candidate HUDF12-4106-7304 of \citet{Ellis13} 
   \textit{(left)} next to an image of a diffraction spike of a nearby star \textit{(center)}, with the rescaled flux of the diffraction spike subtracted from the UDF12-4106-7304 image \textit{(right)}. The latter was derived by fitting the core of the bright galaxy to the right (West) of HUDF12-4106-7304.
This was done for both F140W and F160W.  The images are centered at the
same pixel offset from the nearby bright object causing the diffraction
spike. The bright object lies 5 arcsec to the right (W) in each case. The
location of the putative $z \sim 9.5$ candidate is marked as red circle. It
clearly lies extremely close to where the peak flux of the diffraction
spike is seen. A hint of a linear spike is seen in the  $JH_{140}$ image, running
across the source location.  While a source is still seen in the subtracted image, its
estimated flux is reduced by a factor $\sim2$, making this a $2.8\sigma$ detection
only. This very weak detection evidence, together with the near-blending of
the source with another foreground galaxy, strongly suggests that this is
not a real source. Such faint higher-order diffraction spikes are a
well-known pitfall when pushing the data to its limits, in particular in
ultra-deep imaging data, which is mostly taken at the same rotation angle. }
	\label{fig:spike}
\end{figure}

\subsection{Comparison to Previous $z>8$ Samples}

In the following sections, we compare our LBG samples with previous selections over these fields in the literature.

\subsubsection{Bouwens et al. (2011) $z\gtrsim8.5$ $Y$-dropouts}


In \citet{Bouwens11a}, we already identified three possible sources at $8.5\lesssim z\lesssim 10$. These sources were identified based on their red $Y_{105}-J_{125}$ colors. With the advent of the HUDF12 data, all these source are confirmed as valid high-redshift candidates. However, they are all weakly detected in the $Y_{105}$ filter, which results in a somewhat lower estimate on their redshift. Nevertheless, one of these sources (UDFy-38135540) is included in our present $z\sim9$ sample. The other two (UDFy-37796000 and UDFy-33436598) do have $YJ-JH$ colors of $\sim0.4$, which are too blue to be included in our sample. Their photometric redshifts are 7.8 and 7.7, respectively. The photometry of both these sources are also listed in the appendix in Table \ref{tab:additional}.

As deeper data becomes available it is not unusual to find that the
photometric redshifts undergo small shifts to lower values, also due to the larger number of sources at lower redshifts \citep{Munoz08}.  The original
bias to higher redshifts results from the larger photometric scatter in
shallower data, resulting in an overestimate of the Lyman break amplitude.
Similar biases also affect the photometric redshift samples, e.g., compare
the redshift estimates of \citet{McLure10} with \citet{McLure12}.   This is
a well-known and well-understood effect and should be expected to affect
all redshift estimates derived from photometric data, regardless of the
procedures used. This effect explains the  question raised by \citet{Ellis13} regarding the slightly lower
redshift for these sources.  


\subsubsection{Bouwens et al. (2011) $z\sim10$ Candidates}

In our previous analysis of the full HUDF09 data over the HUDF, we already
identified the source XDFjh-39546284 as a probable high-redshift source.
Based on those data and on a plausible evolution of the UV LF to higher
redshift, we expected this source to lie at $z\sim10.4$. Surprisingly however,
XDFjh-39546284 was not detected in the new $JH_{140}$ data
and so its redshift cannot be  $z\sim10.4$ \citep[see also][]{Ellis13}.  Its nature is now unclear. The
best-fit $z=11.8$ solution is quite unlikely, given what we now know 
about the evolution of the LF at redshifts $z\sim 4 - 9$.  XDFjh-39546284 is
$\sim20\times$ brighter than expected for a $z\sim12$ galaxy at its number density \citep[see Figure 4 of][]{Bouwens12c}.  
Dramatic changes to higher luminosity densities at $z>11$ are unlikely 
and so this object presents us with an interesting conundrum. This is
discussed in detail in \citet{Bouwens12c} and \citet{Brammer13}.

After the first half of the HUDF09 data was taken over the HUDF in the
first year of observations, we had identified three potential $z\sim10$
sources \citep[see the Supplementary Information/Appendix A of][]{Bouwens11a}. These sources were selected
as $J_{125}$-dropouts, very similar to the candidates selected in the
present analysis. However, the three candidates were not detected at
sufficient significance in the second year WFC3/IR $H_{160}$ data, which
raised the possibility that they were spurious detections.

With the advent of additional $H_{160}$ data from the HUDF12 survey, we can
now confirm that all these three sources are in fact real. They are all
significantly detected in the full $H_{160}$ and $JH_{140}$ data. However,
only one of these sources is now in our $z>8$ galaxy sample. Two
sources show photometric redshifts of $z\sim8$, given their very faint
detections in the $Y_{105}$ data of 0.5$\sigma$ and 2.3$\sigma$, respectively. 
However, we remark that one of these two
sources may still be at $z>8.5$ given the tentative 
nature of its $Y_{105}$ band detection (i.e. $0.5\sigma$).
The last source (XDFj-38126243)
remains in our new $z\sim10$ $J_{125}$-dropout sample. For this source, we
find a photometric redshift of $z=9.8^{+0.6}_{-0.6}$. With the possible exception of
the enigmatic $z\sim11.8$ redshift candidate (XDFjh-39546284), this is
therefore the highest redshift galaxy candidate in the HUDF12/XDF field.

In Figure \ref{fig:epochStamps}, we show the $JH_{140}$ and $H_{160}$
stamps of the source XDFj-38126243, including splits of the data by epoch. It is clear
that the source is real, as it is now detected at $5.8\sigma$ in $H_{160}$
and at 3.4$\sigma$ in $JH_{140}$. As can be seen, the second year HUDF09
data (Epoch 2), only contains a weak, though statistically-consistent,
signal of this source. 
Given that there were just two epochs available at that time and the overall S/N of 
the source was below our threshold, we did not include this source in the \citet{Bouwens11a} 
and \citet{Oesch12a} analyses.


Also shown in Figure \ref{fig:epochStamps} is the best-fit template and
photometric redshift distribution for this source. With both photometric
redshift codes ZEBRA \citep{Feldmann06,Oesch10c} and EAZY
\citep{Brammer07}, we find a consistent best-fit photometric redshift at
$z=9.8-9.9$ with uncertainties of $\Delta z \sim 0.6$ (see Table
\ref{tab:phot}). As expected for such a faint source, the redshift
likelihood function shows a lower redshift peak around $z\sim2.5$ (gray
SED). The integrated low-redshift ($z<5$) likelihood is 18\%. This is
consistent with our estimate of lower redshift contamination due to
photometric scatter in our $J_{125}$-dropout sample.  Taken together the
data are consistent with this being a viable and likely $z\sim10$
candidate.

Note that \citet{Ellis13} do not include this source in their analysis, 
but they specifically discuss it. They state the source is not significantly
detected ($<5\sigma$) in their summed $J_{125}+JH_{140}+H_{160}$ image, 
for which we see two main reasons. 
(1) 
With a photometric redshift of $z\sim10$, this galaxy is largely redshifted out 
of the $J_{125}$ band, greatly reducing (by $\sim30\%$)
its detection significance in a $J_{125}+JH_{140}+H_{160}$ stack.   Use of a 
$JH_{140}+H_{160}$ stack is better for such cases and is what we do here.
(2) The source is very compact, and therefore it is detected at higher
significance in the small apertures we use here for S/N measurements (0\farcs35 diameter) compared to the
\citet{Ellis13} analysis ($0.47-0.50$\arcsec diameter). Moreover, we stress that  this source is
significantly detected in several independent sub-sets of the data, and is
therefore certainly real (see Figure \ref{fig:epochStamps}).

\subsubsection{Ellis et al. (2013) and HUDF12 Team Papers}

The HUDF12 team has recently published a sample of seven $z\gtrsim8.5$
galaxy candidates identified in the HUDF12 data in \citet{Ellis13}. These
sources were based on a photometric redshift selection technique \citep[see
also][]{McLure12}, with a $>5\sigma$ detection in the
$J_{125}$+$JH_{140}$+$H_{160}$ summed image. However, as shown in Figure
\ref{fig:filters} galaxies start to disappear in $J_{125}$ at $z>9$, which
is why we could expect to find additional sources in our sample compared to
\citet{Ellis13}. Furthermore, we use smaller apertures for S/N measurements than \citet{Ellis13}, which in most cases are 
more optimal for such very small $z\sim9-10$ sources. This results in a small additional gain of $\sim20\%$ in S/N. 

In general, our sample is in very good agreement with the selection of
\citet{Ellis13}. With the exception of two, we include all their sources in our $z>8$ samples.
The discrepant ones are UDF12-3895-7114 and UDF12-4106-7304, which we discuss below. Their photometry is additionally listed in the appendix in Table \ref{tab:additional}.



\textit{UDF12-3895-7114: } This source certainly show colors very similar
to a $z>8$ candidate. However, we measure $(YJ) - JH_{140} = 0.5\pm0.5$,
which is bluer than our selection color for the $z\sim9$ sample. Hence it
is not included in our $z\sim9$ sample.  While \citet{Ellis13} find a
best-fit photometric redshift of $8.6\pm0.7$, the source is not present in
the `robust' sample of \citet{McLure12}, and we find a photometric redshift
distribution function which is very broad, with a best-fit at $z\sim0.5$
(using both ZEBRA or EAZY). 
This different result compared to the \citet{Ellis13} redshift estimate may be caused
by small uncertainties in the photometry measurements (given that we also use different apertures).
Additionally, we perform IRAC flux measurements on a source by source basis. These include
an additional uncertainty due to the subtraction of neighboring sources \citep[see e.g.][]{Labbe10a,Labbe12}.
This is not the case in the \citet{Ellis13} analysis, who note
that they use constant upper limits on the IRAC fluxes.

As we have discussed, photometric redshifts are
very uncertain for sources this faint and so there is a chance that this source is still at $z>8$.
Nevertheless, our analysis raises significant doubt about its high-redshift
nature.

\textit{UDF12-4106-7304: } The WFC3/IR
PSF shows significant diffraction spikes, which are caused by the mount of
the secondary mirror. While typically only seen around bright stars, these
diffraction spikes are so strong that they can also emanate from
compact foreground galaxies, particularly in the redder WFC3/IR filters.
The source UDF12-4106-7304 of \citet{Ellis13} is located at the edge of
such a diffraction spike for both the $H_{160}$ and the $JH_{140}$ filters
(the only filters wherein UDF12-4106-7304 is significantly detected). This
is shown in Figure \ref{fig:spike}. The photometry of this source is
clearly significantly enhanced by the diffraction spike. The detection
significance of UDF12-4106-7304 is critically reduced once the diffraction
spike signal is removed.  The profile of the bright foreground galaxy is
non-trivial to model, but fortunately only its core is relevant for
causing the diffraction spikes. We therefore use galfit \citep{Peng02} to model the center
of this source and subtract the diffraction spikes that were scaled to
match the core flux.  Doing so results in the flux of UDF12-4106-7304 being
reduced by a factor $\sim2\times$, both in $JH_{140}$ and $H_{160}$, which
makes it only a 2.8$\sigma$ total NIR detection. This is too low to be
included in a robust sample.

Additional uncertainty about the reality of this source arises due to its
different morphology in the $H_{160}$ and $JH_{140}$ images.  The `source'
also lies close to another faint foreground galaxy. It is therefore not
clear whether the diffraction spike and the neighboring galaxy conspired
to lead to the detection of this potential candidate.  In any case,
for these reasons, and for the low detection flux, the reality of
UDF12-4106-7304 remains in question and we do not include this source in
our analysis.

\begin{figure}[tbp]
	\centering
	\includegraphics[width=\linewidth]{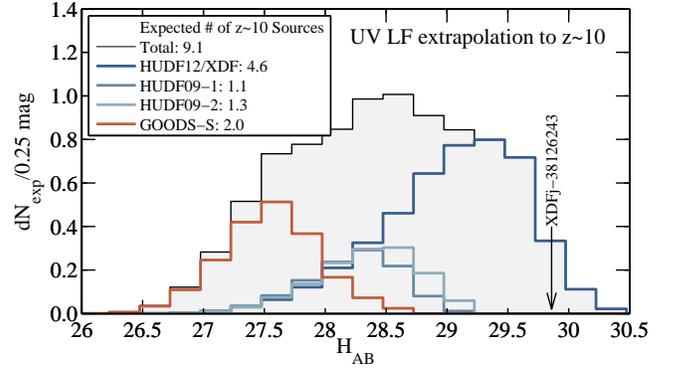}  
  \caption{Expected number of $z\sim10$ candidates per bin of 0.25 mag in the different fields considered in our analysis, assuming that the UV LF evolves steadily from $z\sim8$ to $z\sim10$, consistent with the well-established
trends from $z \sim 8$ to $z \sim 4$.  With this assumption the HUDF12/XDF alone
should have contained 4.6 candidate galaxies at $z \sim 10$. In our whole survey
area, we would have expected to see $\sim9$ sources now. Given that only one
candidate galaxy could be identified (shown by the arrow), this provides strong, direct
evidence that the UV luminosity function and LD are evolving rapidly
from $z\sim10$ to $z\sim8$.}
	\label{fig:Nexp}
\end{figure}

\section{Results}
\label{sec:results}

\subsection{The Abundance of $z>8$ Galaxies}
\label{sec:z8abundance}

The sample of nine $z>8$ galaxy candidates we compiled in the previous sections allows us to make some of the first estimates of the $z\sim9-11$ UV LFs. Although limited in area, the HUDF12/XDF data alone provide very useful constraints already at $z\sim9$ and limits at $z\sim11$. Additionally, due to the deeper data over the HUDF and the CANDELS GOODS-S field compared to our previous analysis in \citet{Oesch12a}, we are able to improve our constraints on the $z\sim10$ LF.

These new constraints at $z\sim9-11$ will allow us to test whether the galaxy population underwent  accelerated evolution at $z>8$ as previously found in \citet{Bouwens11a} and \citet{Oesch12a}, or whether the UV LF trends from lower redshift continue unchanged to $z>8$ \citep[the preferred interpretation of, e.g.,][]{Ellis13,Coe13,Zheng12}. 

In order to test for such accelerated evolution, we start by estimating the  number of galaxies we would have expected to see in our $z\sim9-11$ LBG samples, if the lower redshift trends were to hold unchanged at $z>8$. 
By comparing the observed number of sources with those expected from the
extrapolations we derive a direct estimate of any changes in the evolution
of galaxies at $z>8$.

To derive the expected numbers we use our estimates of the selection
function and completeness measurements described in section
\ref{sec:selfun}. This allows us to compute the number of sources expected
as a function of observed magnitude. For an assumed LF $\phi(M)$, this is
given by:

\begin{equation}
N^\mathrm{exp}_i  = \int_{\Delta m} dm \int dz \frac{dV}{dz} S(m,z)C(m)\phi(M[m,z])
\end{equation}

For the UV LF evolution, we adopt the relations
of \citet{Bouwens11c}: $\phi* = 1.14\times10^{-3} $Mpc$^{-3}$mag$^{-1} = $const,  
$\alpha = -1.73 = $const and $M*(z) = -20.29 + 0.33 \times (z - 6)$.
Note that we assume constant values for the faint-end slope $\alpha$ and the normalization $\phi_*$.
These relations are used as a baseline, when extrapolated to higher redshifts, to
test whether the observed galaxy population at $z>8$ is consistent with the
trends at later times, i.e., at lower redshifts.

With these assumptions, we find that we would
expect to detect a total of $17\pm4$ galaxies in our $z\sim9$ $YJ$-dropout
sample over the HUDF12/XDF field alone. Yet, after correcting for one potential contaminant (see Section \ref{sec:scattersim}), we only detect six sources.
This is $2.8\times$ fewer than expected from the trends from the lower
redshift LFs.

\begin{figure}[tbp]
	\centering
	\includegraphics[width=\linewidth]{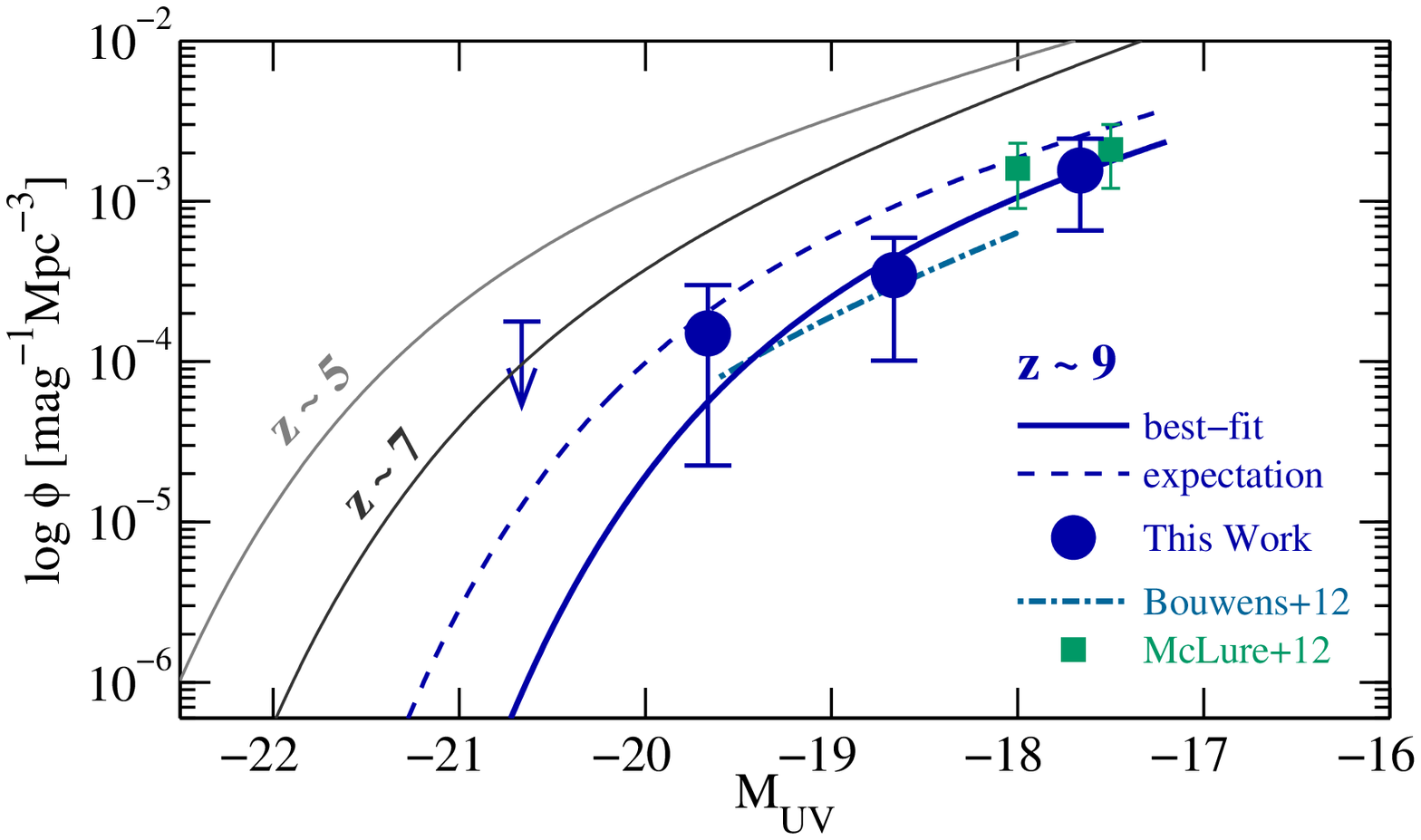}  
	\includegraphics[width=\linewidth]{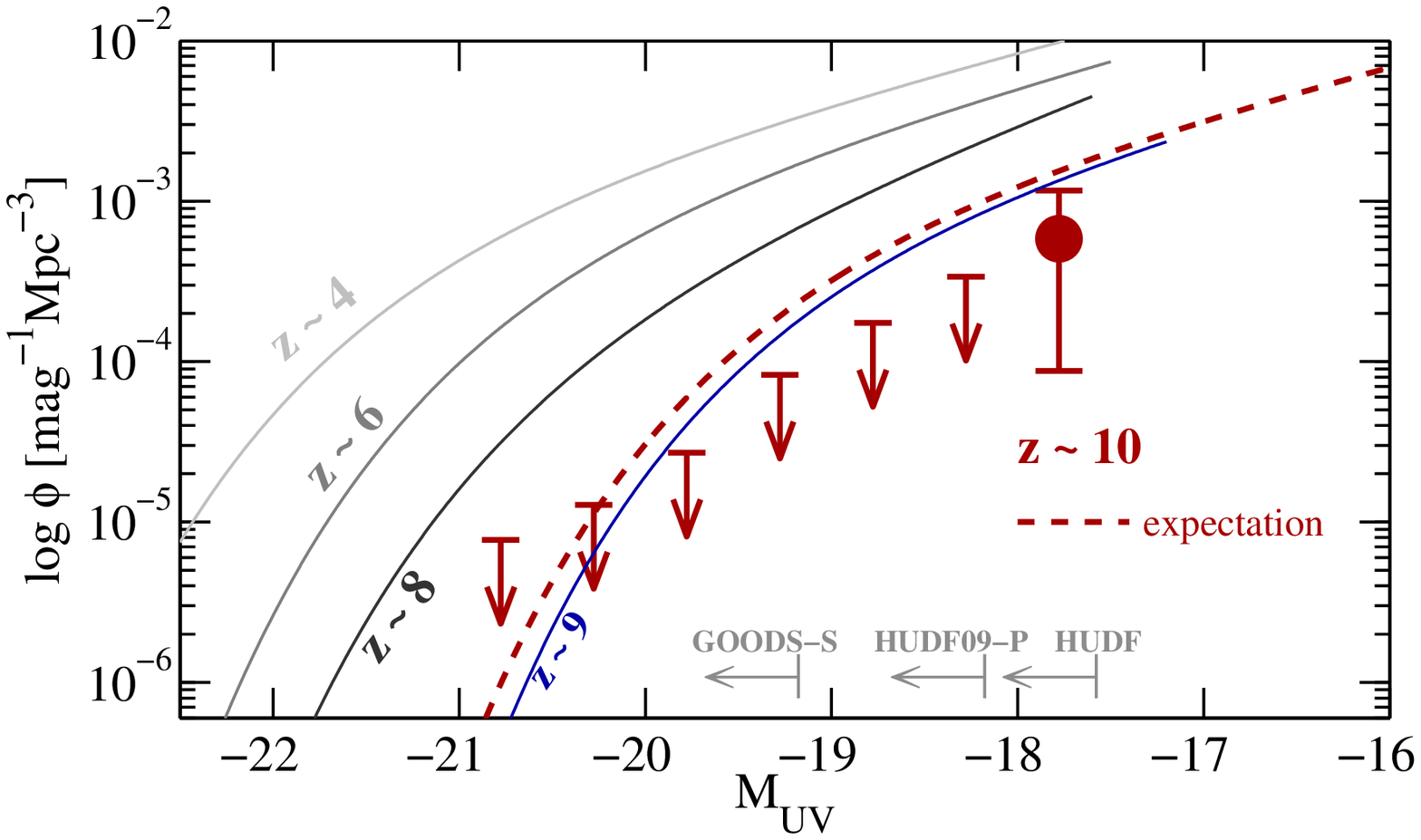}  
  \caption{Constraints on the $z\sim9$ (top) and $z\sim10$ (bottom) UV LF from the HUDF12/XDF data as well as from the additional fields for $z\sim10$. Lower redshift LFs are shown as gray
solid lines for illustration of the LF evolution trends. These are the
most recent determinations from \citet{Bouwens07,Bouwens12b} at $z=4-7$ and \citet{Oesch12b} at $z\sim8$.   
\textit{ (Top) -- } Our step-wise $z\sim9$ LF (dark blue circles) is
computed in bins of 1 mag, which contain 1, 2 and 3 sources, respectively.
These measurements are consistent with (but consistently below) the expected LF
given an extrapolation from lower redshift (dashed blue line).  The
best-fit LF based on luminosity evolution is shown as a solid blue line. This is
derived from the expected number of sources in bins one magnitude wide,
and is a factor $\sim1.5-4\times$ below the extrapolated LF. Also shown as green squares
is the step-wise $z\sim9$ determination from \citet{McLure12} who use a photometric redshift sample derived from HUDF12 data. 
Their determination is in very good agreement with our measurement, although it is unclear why they only constrain the LF at the very faint end. The dot-dashed line represents the best $z\sim9$ LF estimate of \citet{Bouwens12CLASH} over the magnitude range where it is constrained by their lensed candidates from the CLASH dataset. Their determination is based on scaling the normalization of the $z\sim8$ LF to account for the low number density of $z\sim9$ LBG candidates found over the first 19 clusters. Within the current uncertainties, all three determinations of the $z\sim9$ LFs are in good agreement, finding accelerated evolution compared to the lower redshift trends.
\textit{(Bottom) -- } At $z\sim10$, our
analysis includes several additional fields, which is why we can probe to much
lower volume densities than for our $z\sim9$ LF. Nevertheless, since
we only find one potential $z\sim10$ galaxy candidate in our data, we can
mostly only infer upper limits on the LF.  Again, these are consistently
below the extrapolated LF (dashed red line), indicating that the galaxy
population evolves more rapidly at $z > 8$ than at lower redshift. }
	\label{fig:LFevol}
\end{figure}

Similarly, with the same assumptions about the evolution of the UV LF from
$z\sim4$ to $z\sim10$, we would expect to see a total of $9\pm3$ sources in
our $z\sim10$ $J$-dropout sample. We only find one such source, which
suggests that beyond $z\sim9$, the decrement compared to the baseline
evolution is even larger than we found previously from the HUDF09 and the
6-epoch CANDELS data.  We now expect to
see three more sources compared to the six that we expected to see in the
earlier analysis of \citet{Oesch12a}. Yet, no additional  $z\sim10$ sources
are found. 

The expected magnitude distribution for the $z\sim10$
sample is shown in Figure \ref{fig:Nexp}. As indicated in the figure, our
search for $z\sim10$ sources in the CANDELS and ERS fields of GOODS-S
should have resulted in two detections. Only 0.5 sources were expected in
these fields from our previous analysis using somewhat shallower data \citep{Oesch12a}.

Depending on the assumptions about halo occupation, the expected cosmic
variance for a single WFC3/IR field is $40-45\%$
\citep[][]{Trenti08,Robertson10a}. In order to estimate the significance of
our finding of a large offset (i.e., decrement) between the expected number
of sources and that seen, we have to combine the Poissonian and
cosmic variance uncertainties. We estimate the chance of finding $\leq1$
source in our full search area using the appropriate expected number counts
and cosmic variance estimates for the individual search fields. The latter
are based on the cosmic variance calculator of \citet{Trenti08}. Using
simple Monte-Carlo simulations, we derive that given that we expected to find
9 sources, finding $\leq1$ occurs at a probability of only 0.5\%.
Therefore, our data are inconsistent with a simple extrapolation of the
lower redshift LF evolution at 99.5\%.   

This new estimate reinforces the conclusion of \citet{Bouwens11a} and
\citet{Oesch12a} that the evolution in the number density of star forming
galaxies between $z\sim10$ and  $z\sim8$ is large, and larger than expected
from the rate of increase at later times, i.e., the evolution was
accelerated in the $\sim$ 200 Myr from  $z\sim10$ to  $z\sim8$.


%

\begin{deluxetable}{ccc}
\tablecaption{Stepwise Determination of the $z\sim9$ and $z\sim10$ UV LF Based on the Present Dataset \label{tab:z910lf}}
\tablewidth{0 pt}
\tablecolumns{2}
\tablehead{$M_{UV}$ [mag] & $\phi_*$  [10$^{-3}$Mpc$^{-3}$mag$^{-1}$]   }

\startdata
\cutinhead{$z\sim9$}


$-20.66$  &  $<0.18$  \\ 
$-19.66$  &  $0.15^{+0.15}_{-0.13}$  \\ 
$-18.66$  &  $0.35^{+0.24}_{-0.24}$  \\ 
$-17.66$  &  $1.6^{+0.9}_{-0.9}$  \\[0.1cm]

\cutinhead{$z\sim10$}

$-20.78$  &  $<0.0077$\tablenotemark{*}  \\ 
$-20.28$  &  $<0.013$  \\ 
$-19.78$  &  $<0.027$  \\ 
$-19.28$  &  $<0.083$  \\ 
$-18.78$  &  $<0.17$  \\ 
$-18.28$  &  $<0.34$  \\ 
$-17.78$  &  $0.58^{+0.58}_{-0.50}$  

\enddata

\tablenotetext{*}{1$\sigma$ upper limit for a non-detection.}

\end{deluxetable}

\begin{deluxetable*}{lcccccccc}
\tablecaption{Summary and Comparison of $z\gtrsim8.5$ LF Determinations in the Literature \label{tab:lfcomparison}}
\tablewidth{0 pt}
\tablecolumns{5}
\tablehead{Reference & Redshift & $\log\phi_*$  [Mpc$^{-3}$mag$^{-1}$]  &  $M_{UV}^*$ [mag]  &  $\alpha$  }

\startdata
This Work  				&  11  & $-2.94$ (fixed)  & $>18.4$ (1$\sigma$)  & $-1.73$ (fixed) \\
This Work  				&  10  & $-2.94$ (fixed)  & $-17.7\pm0.7$  & $-1.73$ (fixed) \\
This Work  				&  9    &  $-2.94$ (fixed)  & $-18.8\pm0.3$  & $-1.73$ (fixed) \\
 \noalign{\vskip .7ex} \hline \noalign{\vskip 1ex}
\citet{Oesch12a}      &   10  & $-2.96$ (fixed)  & $-18.0\pm0.5$  & $-1.74$ (fixed) \\
\citet{Bouwens12CLASH}  & 9.2 &  $-3.96\pm0.48$  & $-20.04$ (fixed)  &  $-2.06$ (fixed)


\enddata
\end{deluxetable*}

\subsection{The UV Luminosity Function at $z>8$}

The above calculations of the expected number of sources can be used
directly to constrain the UV LFs at $z\sim9-11$.  Since the number of candidates are
small, we need to make some assumptions about how to characterize the
evolution.  The UV LFs at later times provide a valuable guide.  The
parameter that evolves the most is the characteristic luminosity $L*$; the
normalization ($\phi*$ and the faint-end slope ($\alpha$) are relatively
unchanged from $z\sim8$ to  $z\sim4$ (although we do have evidence for 
evolution toward steeper faint-end slopes). This suggests that we should estimate
what evolution in the characteristic luminosity best reproduces the
observed number of sources, while keeping both the normalization and the
faint-end slope fixed.

Doing so results in a best-estimate of the luminosity evolution
of $dM/dz = 0.49\pm0.09$ from $z\sim6$ to $z\sim9$ and $dM/dz =
0.6\pm0.2$ to $z\sim10$. The characteristic magnitudes at these redshifts
are thus expected to be $M_*(z\sim9) = -18.8\pm0.3$ and $M_*(z\sim10) =
-17.7\pm0.7$.  The uncertainties on these measurements are still quite
large, given the small sample sizes and the small area probed, in
particular for the $z\sim9$ search. 

We perform the same calculation for the $z\sim10.7$ $JH_{140}$-dropout sample. 
However, given the uncertain nature of the single candidate
source in that sample (XDFjh-39546284), we treat the estimate at $z\sim10.7$
as an upper limit. We therefore compute the evolution in $M_*$ which is
needed to produce one source or fewer in the sample. This is found to be
$dM/dz>0.4$, which is a less stringent constraint than that for our
$z\sim10$ estimate, due to the much smaller area probed by our $z\sim10.7$
search. The inferred constraint on the characteristic magnitude is
$M_*(z=10.7)>-18.4$ mag. All our estimates of the UV LFs are summarized in Table \ref{tab:lfcomparison}.

%
%

The above estimates for the characteristic luminosity of the UV LF can also be
compared with the characteristic luminosity from the step-wise
determination of the LF using the observed galaxies and limits.  
The step-wise luminosity function is derived using an approximation of the
effective selection volume as a function of observed magnitude $V_{\rm
eff}(m) = \int_0^\infty dz \frac{dV}{dz} S(z,m)C(m)$. The LF is then simply
$\phi(M_i)dM = N^{\rm obs}_i/V_{\rm eff}(m_i)$. 

This derivation is only valid as long as the absolute magnitude varies
slowly with observed magnitude. This is not the case for $z>11$, since the
IGM absorption affects the $H_{160}$ band such that the relation between
luminosity and the observed magnitude becomes strongly redshift dependent.
We therefore restrict our analysis of the step-wise LFs to the $z<10.7$ samples. 
In any case, the small area probed by the HUDF12/XDF
data does currently not significantly constrain the UV LF at $z\sim11$. The step-wise $z\sim9$ and $z\sim10$ LFs are tabulated in Table \ref{tab:z910lf}.

Figure \ref{fig:LFevol} shows our constraints on the UV LF at $z\sim9$ and $z\sim10$. 
The expected UV LFs extrapolated from the lower-redshift
trends are shown as dashed lines. Clearly, at both redshifts, the observed
LF lies significantly below this extrapolation, as expected from our
analysis of the observed number of sources in the previous section. 

The best-fit $z\sim9$ LF using $M_*$ evolution is a factor
$1.5-5\times$ below the extrapolated LF at $M_{UV}>-20$ mag. At the bright
end, the small area probed by the single HUDF12/XDF field limits our LF
constraints to $\gtrsim10^{-4}$ mag$^{-1}$Mpc$^{-3}$, which is clearly too
high to be meaningful at $M_{UV}<-20$.
It will therefore be very important to cover several fields with F140W imaging in the future
in order to push the $z\sim9$ selection volumes to interesting limits.

At $z\sim10$, the use of deeper data both on the HUDF and on the GOODS-S
field allows us to push our previous constraints on the UV LF to fainter
limits. Since we only detect one $z\sim10$ galaxy candidate, however, our constraints
mainly consist of upper limits.  Nevertheless, it is evident that at all
magnitudes $M_{UV}>-20$ mag these upper limits are consistently below the
extrapolated UV LF, up to a factor $4\times$ lower. Additionally, the limits are also clearly below the best-fit $z\sim9$ LF, showing that the
UV LF continues to decline at $z>9$.


\begin{figure}[tbp]
	\centering
	\includegraphics[width=\linewidth]{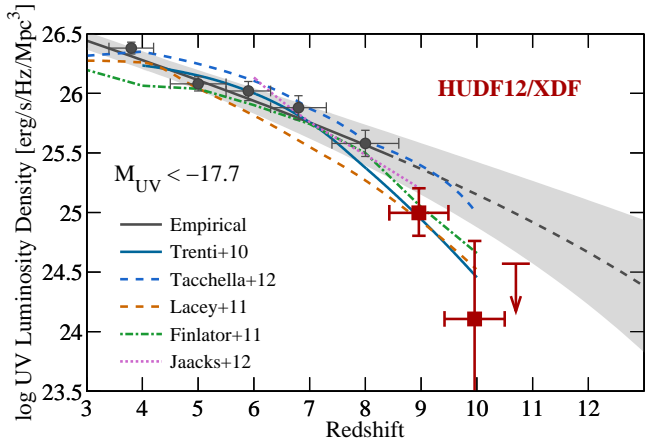}  
  \caption{The evolution of the UV luminosity density (LD) $\rho_{UV}$  contributed
by all galaxies brighter than $M_{1400}=-17.7$ mag. Our new measurements
from the HUDF12/XDF data are shown as red squares. Measurements of the LD
at $z\leq8$ are derived from the UV LFs from \citet{Bouwens07,Bouwens12b}.
No correction for dust extinction has been applied.  The
measurements are plotted at the mode of the redshift distributions shown in
Figure 6. For the highest redshift $z\sim10.7$ $JH$-dropout sample, we only
show an upper limit given that the single source we find in this sample is
either at an even higher redshift (where the selection volume of our data
is very small) or is a low redshift extreme line emitter. The dark gray
line and shaded area represent an extrapolation of the redshift evolution
trends of the $z\sim4-8$ UV LF.  Our LD estimates at $z\sim9-10$ are clearly
lower than this extrapolation. However, the observed rapid build-up of
galaxies at $z \sim 10$ to $z \sim 8$ is not unexpected, since it is consistent with a
whole suite of theoretical models. Some of these are shown as colored lines. They
are halo occupation models \citep[blue solid and dashed,][]{Trenti10,Tacchella12}, a semi-analytical model \citep[orange dashes,][]{Lacey11}, and two hydrodynamical simulations \citep[][]{Finlator11b,Jaacks12a}.
These different models uniformly predict a steepening in the LD evolution
at $z>8$.  The conclusion to be drawn is that the shape of the trend from
$z\sim10$ to $z\sim7$ is mainly due to the rapid build-up of the underlying dark-matter halo
mass function, rather than any physical changes in the star-formation
properties of galaxies.
   }
	\label{fig:LDevol}
\end{figure}

\subsection{The UV Luminosity Density Evolution at $z>8$}

The evolution of the UV luminosity density (LD) at $z>8$ has received
considerable attention in recent papers, triggered by our initial finding
of a significant drop in the LD from $z\sim8$ to $z\sim10$ (i.e., a rapid
increase in the LD within a short period of time).  With the new
HUDF12/XDF data, it is now possible to refine this measurement by adding a
$z\sim9$ and a $z\sim11$ point, while also allowing us to improve upon our
previous measurements at $z\sim10$. 

The UV LDs inferred from our $z>8$ galaxy samples are shown in Figure
\ref{fig:LDevol}. The measurements show the LD derived by integrating the
best-fit UV LF determined in the previous section.   The integration limit is
set to $M_{UV} = -17.7$ mag, which is the current limit probed by the
HUDF12/XDF data.  For comparison, we also show the lower redshift LD
measurements from the compilation of \citet{Bouwens07,Bouwens12c}. These were computed in the
same manner as the new $z>8$ values, and were not corrected for dust extinction. 
A summary of our measurements for the LD are listed in Table
\ref{tab:LDSFRD}.

As can be seen, our new measurements at $z>8$ lie significantly below the
$z\sim8$ value. The decrement in LD from $z\sim8$ to $z\sim9$ is
$0.6\pm0.2$ dex, and it is even larger at $1.5\pm0.7$ dex to $z\sim10$.
Therefore, our data confirms our previous finding of more than an order of
magnitude increase of the UV LD in the short time period, only 170 Myr,
from $z\sim10$ to $z\sim8$.

The gray line and shaded area show the expected LD evolution when
extrapolating the $z\sim4-8$ Schechter function trends to higher
redshift. All our measurements at $z>8$ lie below the extrapolation.
Although the offsets individually are not large (they are $<2\sigma$), the
consistent offset to lower LD supports a hypothesis that significant changes are
occurring in the LD evolution at $z>8$.

It is interesting to note that this offset to lower LD is not unexpected,
as it is also seen in several theoretical models.  In Figure
\ref{fig:LDevol}, we compare our observational results to two conditional luminosity function models from
\citet{Trenti10} and \citet{Tacchella12}, the prediction from a semi-analytical model of
\citet{Lacey11}, as well as the results from two hydrodynamical simulations of
\citet{Finlator11b} and \citet{Jaacks12a}.  

All these models are in relatively good agreement with the lower redshift
($z<8$) measurements. As can be seen from the figure, essentially all
models do show a steeper evolution at $z>8$ than a purely empirical
extrapolation of the UV LF further into the epoch of reionization. Since
these models are all very different in nature, this strongly suggests that
the rapid build-up we observe in the galaxy population is mainly driven by
the build-up in the underlying dark matter halo mass function, which is
also evolving very rapidly at these epochs.

\begin{deluxetable*}{cccc}
\tablecaption{Summary of Luminosity Density and Star-Formation Rate Density Estimates\label{tab:LDSFRD}}
\tablewidth{0 pt}
\tablecolumns{5}
\tablehead{\colhead{Dropout Sample} & \colhead{Redshift} & \colhead{$\log_{10}\rho_{UV}$\tablenotemark{$\dagger$} } &\colhead{$\log_{10}\rho_{*}$}  \\
 &  &$[$erg$~$s$^{-1}$Hz$^{-1}$Mpc$^{-3}]$ & $[$\msol yr$^{-1}$Mpc$^{-3}]$
}

\startdata
YJ  &  9.0  &  $25.00^{+0.19}_{-0.21}$ & $-2.86^{+0.19}_{-0.21}$ \\ 
J  &  10.0  &  $24.1^{+0.7}_{-0.9}$ & $-3.7^{+0.7}_{-0.9}$ \\ 
JH & 10.7 & $<24.6$ & $<-3.3$ \\ \noalign{\vskip .7ex} \hline \noalign{\vskip 1ex}

B &3.8 & $26.38\pm0.05$ & $-1.21 \pm 0.05$ \\
V &5.0& $26.08\pm0.06$ & $-1.54 \pm 0.06$ \\
i &5.9& $26.02\pm0.08$ & $-1.72 \pm 0.08$ \\
z& 6.8& $25.88\pm0.10$ & $-1.90 \pm 0.10$ \\
Y & 8.0 & $25.58\pm0.11$ &  $-2.20 \pm 0.11$ 

\enddata

\tablenotetext{$\dagger$}{Integrated down to $0.05L_{z=3}^*$ ($M_{1400}=-17.7$ mag)}
\tablecomments{The lower redshift data points are based on the UV LFs from \citet{Bouwens07}, \citet{Bouwens11c}, and \citet{Oesch12b} }
%
\end{deluxetable*}

\begin{figure*}[tbp]
	\centering
	\includegraphics[width=0.7\linewidth]{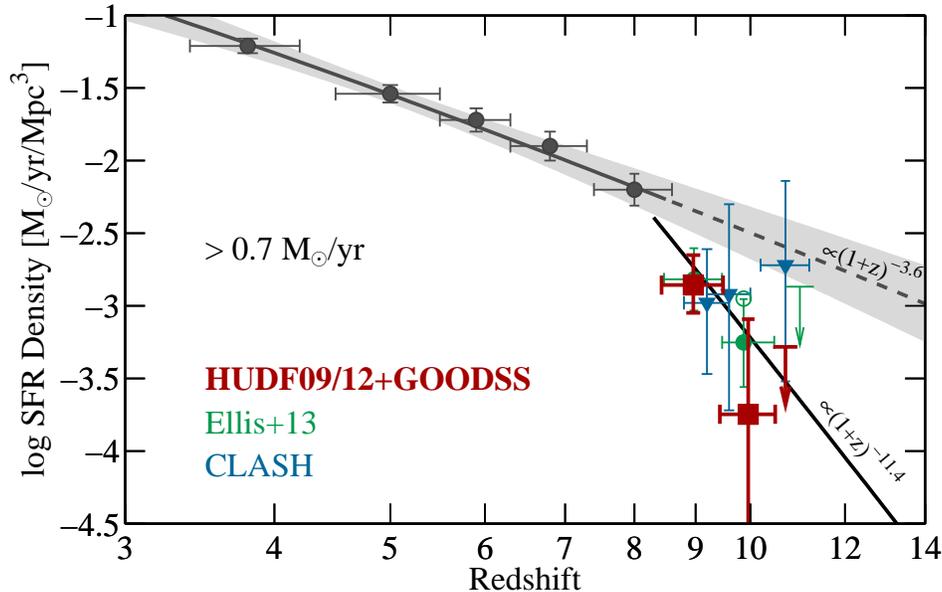}  
  \caption{ The evolution of the star-formation rate density (SFRD)  $\dot\rho_*$
contributed by star-forming galaxies brighter than $M_{1400}=-17.7$ mag.  This
limit corresponds to a star-formation limit $>0.7 M_\odot$yr$^{-1}$. Measurements at
$z>8$ are shown from the present analysis (dark red squares), from the HUDF12
analysis of \citet{Ellis13} (green circles) as well as from CLASH
cluster detections \citep[blue triangles;][]{Bouwens12CLASH,Coe13,Zheng12}.
 The lower redshift datapoints are derived from UV LFs
from \citet{Bouwens07,Bouwens12b}. The SFRD is derived from the UV LD
integrated to these limits and corrected for dust extinction using the most
recent estimates from \citet{Bouwens12a}.  A clear decrement by 0.6
dex in the SFRD is consistently seen between the measurement at $z\sim8$ and
that at $z\sim9$. At redshifts higher than $z\sim9$, all datapoints only contain
one galaxy, resulting in large uncertainties. (We have corrected down
the measurement of Ellis et al.\ to account for a likely diffraction spike source;
c.f.  open vs. filled circle at $z\sim10$.) Given the large
uncertainties, the individual $z>8$ measurements are all consistent with each other.
At $z=4-8$ the SFRD increases gradually, following $\dot\rho_*\propto (1+z)^{-3.6\pm0.3}$
(dark gray line).  The extrapolation of this trend to higher redshift is
shown by the dashed line and gray region (1$\sigma$). All data points lie
below this line, indicating that the extrapolation is not a good fit.  The
combined best-fit evolution using all the CLASH measurements and our new HUDF12/XDF results is
significantly steeper, following $(1+z)^{-11.4\pm3.1}$ (black solid line).
While we find an increase by a factor 30$\times$ in the SFRD between $z\sim10$ and $z\sim8$
from our HUDF12+GOODSS analysis alone, the best-fit trend results in a
somewhat reduced increase. Nevertheless, this is still a factor $10\times$ in the
$\sim 170$ Myr from $z\sim10$ to $z\sim8$, i.e., a large change
over a short time period.
   }
	\label{fig:SFRDevol}
\end{figure*}

\subsection{The SFR Density Build-up from $z\sim11$ to $z\sim8$}

Our results combined with those of others now provide a substantially
larger sample at $z\gtrsim8$ for estimate of the SFR density than was
available for \citet{Bouwens11a} and \citet{Oesch12a}.  The SFR density at
$z>8$ was recently estimated based on four high-redshift galaxies
identified in the CLASH survey \citep{Bouwens12CLASH,Coe13,Zheng12}, and
from seven galaxies identified in the HUDF12 data \citep{Ellis13}. We
present all these results, together with our own measurements in Figure
\ref{fig:SFRDevol}, where we plot the SFR density as a function of redshift
in star-forming galaxies with SFR $ > 0.7 M_\odot$yr$^{-1}$ (corresponding to a magnitude limit of $M_{UV}=-17.7$ mag).

The SFR densities are derived from the UV LD estimates after correction for dust extinction. We use the most
recent determinations of the UV continuum slopes $\beta$ as a function of
UV luminosity and redshift from \citet{Bouwens12a} together with the \citet{Meurer99} $\beta$-extinction relation. The dust-corrected LDs are then
converted to SFR densities using the conversion factor of \citet{Madau98},
assuming a Salpeter initial mass function.

Across $z\sim 4$ to $z\sim8$, the star-formation rate density clearly evolves
very uniformly.  The evolution is well reproduced by a power law
$\dot\rho_*\propto(1+z)^{-3.6\pm0.3}$, which is shown as dark gray line
in Figure \ref{fig:SFRDevol}.  Interestingly, all measurements lie below the extrapolation of this trend to
higher redshift.  Again, individual measurements are within $<2\sigma$ of
the trend, but the offset to smaller SFR densities in the mean is very
clear.

Note that each of the three groups find a consistent decrement of the SFRD from $z\sim8$ to $z\sim9$. 
Specifically, from our data, we find a drop by $0.6\pm0.2$ dex. This is
$\sim2\sigma$ below the simple extrapolation of the lower redshift trends.

At $z\sim9$ our SFRD estimate is in excellent agreement with the
measurement of \citet{Ellis13} based on a photometric redshift selection.
At $z\sim10$, however, we find a significantly lower value, mainly due to
our inclusion of a larger dataset covering a larger area, in which we do
not find any additional candidate.  Nevertheless, our measurement is within
$\sim1\sigma$ of the result of \citet{Ellis13}, in particular, after we
correct their measurement down by a factor two to account for the source
that is likely the result of a diffraction spike in their $z=9.5$
sample (Figure \ref{fig:spike}). 

Also at $z\sim10.7$, our upper limit is significantly below the $z=10.5$ estimate
of \citet{Ellis13}. This is likely due to the wide redshift range
probed by our $JH_{140}$-dropout sample, which extends from $z\sim9.5$ to
$z\sim11.8$ (see Figure \ref{fig:zSelection}). \citet{Ellis13} consider a
strict boundary of $z=10$ to $z=11$.


In order to compute the best-fit evolution of the SFRD at $z\geq8$, we
 combine all $z>8$ measurements from CLASH with our improved estimates at
$z\sim9-11$, together with the previous $z\sim8$ SFRD measurement \citep{Oesch12b}. 
The best-fit evolution falls off very rapidly, following $\dot\rho_* \propto (1+z)^{-11.4\pm3.1}$.
This is shown as the black line in Figure \ref{fig:SFRDevol}. As can be seen, this is significantly
steeper than the $z\sim4-8$ trends. By $z\sim10$, the best-fit evolution is already a factor $\sim5\times$ below the lower redshift trend.  Therefore, the combined constraint on the SFRD evolution
from all datasets in the literature clearly points to an accelerated evolution at $z>8$.

\section{Summary and Conclusions}
\label{sec:summary}

We have used the new, ultra-deep WFC3/IR data over the HUDF field as well as the optical XDF data 
to provide a reliable selection of 
galaxies at $z>8$. The new observations from the HUDF12 program push the depth of the $H_{160}$ imaging
deeper by $\sim$0.2 mag compared to our previous data from the HUDF09 survey,
and they provide additional $JH_{140}$ imaging. The $JH_{140}$ data are
very useful for selecting some of the first $z\sim9$ and $z\sim11$ galaxy samples
using the Lyman Break Technique.  Furthermore, we extended our previous
search for $z\sim10$ galaxies \citep{Oesch12a} to fainter limits by including this new HUDF12/XDF data set. Our analysis is the
most extensive search for $z>8$ galaxies to date.

From our full dataset, we find a total sample of nine $z>8$ galaxy candidates.
Seven of these lie in our $z\sim9$ selection. 
Contamination is always a central concern for high redshift
samples and after careful analysis we expect that the contamination
fraction is small, being only about $15-20\%$.  

We found that one of the $z\sim9$ sources has a very wide photometric redshift likelihood distribution, with an ambiguous best-fit at $z_{phot} = 2.2$ (using ZEBRA; $z_{phot} = 8.6$ using EAZY). We therefore exclude this source from the subsequent analysis.


We discover a new $z\sim10$ source (at $z=9.8\pm0.6$), making it one of the
very few galaxies known at this very high redshift, just 460 Myr after
the Big Bang.
The highest redshift candidate in our sample is XDFjh-39546284 that was
previously identified at $z\sim10.4$. However, these new data (the $JH_{140}$
in particular) constrain this galaxy to be at $z\sim11.8$, if it is at high
redshift. This interpretation is
problematic and has led to discussion about it being a possible
lower-redshift $z\sim2$ object, and so its true nature remains quite
uncertain at this time \citep[see][]{Ellis13,Bouwens12c,Brammer13}.

This sample of $z>8$ galaxy candidates proves to
be very important for setting a number of constraints on galaxy
build-up at very early times, allowing us to derive an estimate of
the UV LF at $z\sim9$, to improve our constraints at $z\sim10$, and to set
limits at $z\sim11$. 

The main result from our analysis is a confirmation of our previous finding
that the galaxy population, as seen down to $M_{UV} = -17.7$ mag, evolves much more rapidly at $z>8$ than
at lower redshift \citep[identified as "accelerated evolution";][]{Bouwens11a,Oesch12a}. This is seen in
(1) the expected number of galaxies when extrapolating the lower redshift
trends to $z>8$ (Figure \ref{fig:Nexp}), (2) in the direct constraints on
the UV LF (Figure \ref{fig:LFevol}), and (3) in the evolution of the
luminosity and star-formation rate densities down to our current completeness limits (Figures \ref{fig:LDevol},
\ref{fig:SFRDevol}). All measurements consistently point to accelerated
evolution at early times, beyond $z\sim8$.

Specifically, if the lower redshift trends of the UV LF are extrapolated to
$z\sim10$, we would have expected to see $9\pm3$ candidate sources in our
full data set, $\sim5$ of which only in the HUDF12/XDF data alone. However,
only one such candidate is found in the HUDF12/XDF data, which suggests that the galaxy population evolves
more rapidly than at lower redshift at 99.5\% significance (see Section \ref{sec:z8abundance}).

From $z\sim8$ to $z\sim9$, the luminosity density in star-forming galaxies
with SFR$ >0.7 M_\odot$yr$^{-1}$ (i.e. $M_{UV}<-17.7$) decreases by $0.6\pm0.2$ dex. This decrement is fully consistent with previous estimates from CLASH \citep{Bouwens12CLASH} and from the HUDF12 data alone \citep{Ellis13}.

The combination of our new measurements of the SFRD at $z>8$ with all previous 
estimates from the CLASH data \citep{Bouwens12CLASH,Coe13,Zheng12} results in a
best-fit evolution which is extremely steep, following $\dot\rho_* \propto (1+z)^{-11.4\pm3.1}$.

These results on the galaxy evolution at $z>8$ contrast with the
conclusions drawn by several recent papers, who argue that the UV LD evolution 
at $z>8$ is consistent with the lower redshift trends \citep[e.g.][]{Ellis13,Coe13,Zheng12}. 
However, the small sample sizes
of $z>8$ galaxies in these individual analyses resulted in large
uncertainties on the LD and SFRD evolution. We show here that once all
these measurements are combined self-consistently, they do indeed point to
accelerated evolution at $z>8$, consistent with theoretical expectations.

Note that the steep fall-off we find in the UV LD at $z>8$ is not at odds with galaxies driving reionization. 
Our measurements only reach to $\sim L*$ at $z\sim10$ (i.e. to $M_{UV} = -17.7$ mag). However, with the
steep faint-end slopes that are consistently found for $z>4$ UV LFs, the total luminosity density is completely dominated
by galaxies below this threshold \citep[see e.g.][]{Bouwens12b,Kuhlen12}.

With WFC3/IR we are now in a similar situation in studying $z\sim9.5-10$
as we were three years ago with NICMOS at $z\sim7$.  Galaxy samples are still
small, and the conclusions are uncertain. However, over the next few years
the $z>8$ frontier will be explored more extensively. In particular, the additional deep field 
observations to be taken as part of the Deep Fields Initiative (a large Director's 
Discretionary program), will significantly increase sample sizes and should allow for
improved constraints on the $z\sim9$ and $z\sim10$ LF at $M_{UV}<-18.5$ mag. 
This will enable more precise constrains on the accelerated evolution that we now see in the galaxy population
from the data over GOODS-South.

\acknowledgments{Support for this work was provided by NASA through Hubble Fellowship grant HF-51278.01 awarded by the Space Telescope Science Institute, which is operated by the Association of Universities for Research in Astronomy, Inc., for NASA, under contract NAS 5-26555.
Additionally, this work has been supported by NASA grant NAG5-7697 and NASA grant HST-GO-11563.01. }

Facilities: \facility{HST(ACS/WFC3), Spitzer(IRAC)}.

\bibliographystyle{apj}


\appendix


\setlength{\tabcolsep}{0.04in} 

\begin{deluxetable}{cccccccccccccl}
\tablecaption{Photometry of $z>8$ LBG Candidates in the HUDF12/XDF Data\label{tab:phot}}
\tablewidth{\linewidth}
\tablecolumns{11}

\tablehead{\colhead{ID} & RA & DEC &\colhead{$H_{160}$}  & \colhead{$(YJ)-JH_{140}$}   &\colhead{$J_{125}-H_{160}$} &   \colhead{$JH_{140}-H_{160}$}  & \colhead{S/N$_{H160}$}  & \colhead{S/N$_{JH140}$}  & \colhead{S/N$_{J125}$} & \colhead{$\chi^2_{opt}$}  \\ 
   &  \colhead{$z_{phot}^{ZEBRA}$}  & \colhead{$z_{phot}^{EAZY}$}  & \colhead{comments}}

\startdata
\cutinhead{$z\sim9$ YJ-dropouts}


XDFyj-38135540  &  03:32:38.13  &  -27:45:54.0  &  $27.95\pm0.10$ & $0.8 \pm 0.1$  & $0.0 \pm 0.1 $  & $-0.2 \pm 0.1$  & 13.1  & 16.0 & 9.8 & 0.2 \\  
               & $8.4^{+0.1}_{-0.1}$               &  $8.4^{+0.1}_{-0.2}$              &  \multicolumn{8}{l}{Bouwens UDFy-38125539;  McLure HUDF12-3813-5540 ($z=8.3$); and in other $Y$-dropout samples.}  \\[0.3cm] 

XDFyj-39478076  &  03:32:39.47  &  -27:48:07.6  &  $28.53\pm0.14$ & $0.8 \pm 0.2$  & $0.4 \pm 0.2 $  & $-0.0 \pm 0.2$  & 8.7  & 9.7 & 5.0 & -0.6 \\  
                 & $8.1^{+0.3}_{-0.6}$               &  $8.3^{+0.2}_{-0.5}$              &  \multicolumn{8}{l}{Bouwens UDFy-39468075; Ellis HUDF12-3947-8076 ($z=8.6$)} \\[0.3cm]

XDFyj-39216322  &  03:32:39.21  &  -27:46:32.2  &  $29.49\pm0.25$ & $1.1 \pm 0.5$  & $0.3 \pm 0.4 $  & $0.1 \pm 0.3$  & 5.1  & 4.9 & 3.2 & -1.4 \\  
                            & $8.8^{+0.5}_{-0.5}$                &  $8.9^{+0.5}_{-0.4}$               &  \multicolumn{8}{l}{Ellis HUDF12-3921-6322 ($z=8.8$)} \\[0.3cm]
                             
XDFyj-42647049  &  03:32:42.64  &  -27:47:04.9  &  $29.15\pm0.21$ & $1.5 \pm 0.7$  & $0.8 \pm 0.6 $  & $0.1 \pm 0.3$  & 4.4  & 4.9 & 2.1 & -0.5 \\  
                            & $9.0^{+0.5}_{-0.5}$                &  $9.2^{+0.5}_{-0.6}$               &  \multicolumn{8}{l}{Ellis HUDF12-4265-7049 ($z=9.5$)} \\[0.3cm]

XDFyj-40248004  &  03:32:40.24  &  -27:48:00.4  &  $29.87\pm0.30$ & $1.3 \pm 0.7$  & $0.3 \pm 0.6 $  & $-0.1 \pm 0.4$  & 3.5  & 3.3 & 2.2 & 0.0 \\  
                        & $8.8^{+0.5}_{-0.5}$               &  $8.9^{+0.6}_{-0.3}$              &  \multicolumn{8}{l}{Faint source. Not in Ellis et al. (2013) sample.} \\[0.3cm]
                             
XDFyj-43456547  &  03:32:43.45  &  -27:46:54.7  &  $29.69\pm0.42$ & $1.3 \pm 0.7$  & $0.5 \pm 0.7 $  & $-0.1 \pm 0.4$  & 3.1  & 3.5 & 1.8 & -3.8 \\  
                         & $8.7^{+0.6}_{-0.5}$               &  $8.9^{+0.7}_{-0.8}$              &  \multicolumn{8}{l}{Ellis HUDF12-4344-6547 ($z=8.8$)} \\[0.3cm]

XDFyj-39446317\tablenotemark{a}  &  03:32:39.44  &  -27:46:31.7  &  $29.77\pm0.27$ & $1.1 \pm 0.7$  & $> 1.0$  & $-0.3 \pm 0.5$  & 3.8  & 3.7 & 1.3 & 1.3 \\  
                          & $2.2^{+0.8}_{-0.7}$               &  $8.6^{+1.0}_{-1.5}$              &  \multicolumn{8}{l}{Faint source. Very wide $p(z)$, with low best-fit redshift. Not in Ellis et al. (2013) sample.} \\[0.4cm]

\cutinhead{$z\sim10$ J-dropouts}
XDFj-38126243  &  03:32:38.12  &  -27:46:24.3  &  $29.87\pm0.40$ & $> 1.9$  & $1.4 \pm 0.9 $  & $0.3 \pm 0.4$  & 5.8  & 3.4 & 1.2 & -0.6 \\  
                & $9.8^{+0.6}_{-0.6}$               &  $9.9^{+0.7}_{-0.6}$              &  \multicolumn{8}{l}{This source was selected as a $z\sim10$ candidate in the HUDF09 year 1 data, but did not appear in } \\
                                &               &              &  \multicolumn{8}{l}{final Bouwens et al. (2011) sample due to low S/N in 2nd year data (see Fig \ref{fig:epochStamps}).} \\[0.4cm]

\cutinhead{$z\sim10.7$ JH-dropouts}
XDFjh-39546284  &  03:32:39.54  &  -27:46:28.4  &  $28.55\pm0.14$ & --  & $> 2.3$  & $> 2.3$  & 7.3  & 0.2 & -1.6 & 0.8 \\  
              & $11.8^{+0.2}_{-0.4}$               &  $11.9^{+0.2}_{-0.5}$               &  \multicolumn{8}{l}{HUDF09 source of Bouwens et al. (2011), Oesch et al. (2012); Ellis HUDF12-3954-6284 ($z=11.9$)} \\ 

\enddata

\tablenotetext{a}{Due to the low photometric redshift estimate, we do not include the source XDFyj-39446317 in our analysis of the UV LF at $z\sim9$. One contaminating lower redshift source is expected in our sample due to photometric scatter (see Section \ref{sec:scattersim}).}
\tablecomments{S/N are measured in circular apertures of fixed 0\farcs35 diameter}

\end{deluxetable}

\begin{deluxetable}{cccccccccccccl}
\tablecaption{Photometry of Additional Potential $z>8$ LBG Candidates not used in this Analysis\label{tab:additional}}
\tablewidth{\linewidth}
\tablecolumns{11}

\tablehead{\colhead{ID} & RA & DEC &\colhead{$H_{160}$}  & \colhead{$(YJ)-JH_{140}$}   &\colhead{$J_{125}-H_{160}$} &   \colhead{$JH_{140}-H_{160}$}  & \colhead{S/N$_{H160}$}  & \colhead{S/N$_{JH140}$}  & \colhead{S/N$_{J125}$} & \colhead{$\chi^2_{opt}$}  \\ 
   &  \colhead{$z_{phot}^{ZEBRA}$}  & \colhead{$z_{phot}^{EAZY}$}  & \colhead{comments}}

\startdata
42126501  &  03:32:42.12  &  -27:46:50.1  &  $28.45\pm0.05$ & $1.7 \pm 0.2$  & $1.4 \pm 0.2 $  & $0.3 \pm 0.1$  & 22.2  & 15.8 & 5.2 & -0.1 \\  
              & $9.5^{+0.1}_{-0.1}$                &  $9.7^{+0.1}_{-0.2}$               &  \multicolumn{8}{l}{This potential source is completely blended with a foreground galaxy.} \\[0.25cm]
              
43246481  &  03:32:43.24  &  -27:46:48.1  &  $28.61\pm0.17$ & $0.8 \pm 0.6$  & $1.3 \pm 0.7 $  & $0.7 \pm 0.4$  & 5.4  & 3.0 & 2.8 & -1.9 \\  
                & $2.5^{+6.1}_{-0.7}$                &  $2.4^{+7.7}_{-0.4}$               &  \multicolumn{8}{l}{Close to bright, clumpy foreground galaxy.} \\[0.25cm]
                
43286481  &  03:32:43.28  &  -27:46:48.1  &  $28.53\pm0.17$ & $> 1.6$  & $> 1.9$  & $0.8 \pm 0.4$  & 6.0  & 2.5 & 0.5 & -0.2  \\ 
             & $10.4^{+0.5}_{-0.5}$               &  $10.6^{+0.6}_{-0.3}$              &  \multicolumn{8}{l}{Close to bright, clumpy foreground galaxy.} \\[0.5cm]

\cutinhead{Additional Sources From Ellis et al. (2013)}
             
UDF12-4106-7304  &  03:32:41.06  &  -27:47:30.4  &    -- & --  & --  & --  & --  & -- & -- & -- \\
           & --   &  --  &   \multicolumn{8}{l}{The photometry of this source is significantly affected by a diffraction spike (see Fig. \ref{fig:spike}).} \\
           &       &        &   \multicolumn{8}{l}{After subtraction of the expected flux of the spike, the source is only a $2.8\sigma$ total NIR detection.} \\[0.25cm]

UDF12-3895-7115  &  03:32:38.95  &  -27:47:11.5  &  $30.02\pm0.30$ & $0.5 \pm 0.5$  & $0.1 \pm 0.5 $  & $0.1 \pm 0.5$  & 3.5  & 3.7 & 3.7 & 1.2 \\  
           & $0.5^{+5.8}_{-0.5}$   &  $0.6^{+8.2}_{-0.5}$  &   \multicolumn{8}{l}{Does not satisfy our color selection: $(YJ)-JH_{140}=0.5<0.75$.}                         \\
             &       &        &   \multicolumn{8}{l}{Additionally, the best-fit photometric redshift is only 0.5 using our photometry.} \\[0.4cm]

\cutinhead{Additional Previous $z\gtrsim8$ Candidates From Bouwens et al.\ (2012)}

%
UDFy-37806001  &  03:32:37.80  &  -27:46:00.1  &  $28.39\pm0.12$ & $0.4 \pm 0.1$  & $-0.2 \pm 0.2 $  & $-0.2 \pm 0.1$  & 10.1  & 13.7 & 9.4 & 2.7 \\  
 & $7.8^{+0.2}_{-0.4}$               &  $7.9^{+0.3}_{-0.4}$              &  \multicolumn{8}{l}{Too blue in $(YJ)-JH_{140}$ for our selection.}  \\[0.25cm] 

UDFy-33446598  &  03:32:33.44  &  -27:46:59.8  &  $29.00\pm0.20$ & $0.3 \pm 0.2$  & $0.1 \pm 0.3 $  & $0.1 \pm 0.2$  & 5.6  & 6.0 & 4.7 & -1.9 \\  
 & $7.7^{+0.4}_{-0.4}$               &  $7.7^{+0.5}_{-0.4}$              &  \multicolumn{8}{l}{Too blue in $(YJ)-JH_{140}$ for our selection}

\enddata

\tablecomments{S/N are measured in circular apertures of fixed 0\farcs35 diameter.}

\end{deluxetable}

\end{document}